\documentclass[fleqn,usenatbib]{mnras}

\usepackage{newtxtext,newtxmath}

\usepackage[T1]{fontenc}

\DeclareRobustCommand{\VAN}[3]{#2}
\let\VANthebibliography\thebibliography
\def\thebibliography{\DeclareRobustCommand{\VAN}[3]{##3}\VANthebibliography}

\usepackage{graphicx}	
\usepackage{amsmath}	

\title[Baryonic impact on WL peak counts]{The FLAMINGO project: Baryonic impact on weak gravitational lensing convergence peak counts}

\author[J. C. Broxterman et al.]{
Jeger C. Broxterman,$^{1,2}$\thanks{E-mail: broxterman@lorentz.leidenuniv.nl}
Matthieu Schaller,$^{1,2}$ 
Joop Schaye,$^{2}$ 
Henk Hoekstra,$^{2}$ 
Konrad Kuijken,$^{2}$ \newauthor
John C. Helly,$^{3}$ 
Roi Kugel,$^{2}$ 
Joey Braspenning,$^{2}$ 
Willem Elbers,$^{3}$ 
Carlos S. Frenk,$^{3}$ 
Juliana Kwan,$^{4}$ \newauthor
Ian G. McCarthy,$^{4}$ 
Jaime Salcido,$^{4}$ 
Marcel P. van Daalen$^{2}$ 
and Bert Vandenbroucke$^{2}$ 
\\
$^{1}$Lorentz Institute for Theoretical Physics, Leiden University, PO Box 9506, NL-2300 RA Leiden, the Netherlands
\\
$^{2}$Leiden Observatory, Leiden University, PO Box 9513, NL-2300 RA Leiden, The Netherlands\\
$^{3}$Institute for Computational Cosmology, Department of Physics, University of Durham, South Road, Durham, DH1 3LE, UK\\
$^{4}$Astrophysics Research Institute, Liverpool John Moores University, Liverpool L3 5RF, UK
}

\date{Accepted XXX. Received YYY; in original form ZZZ}

\pubyear{2023}

\begin{document}
\label{firstpage}
\pagerange{\pageref{firstpage}--\pageref{lastpage}}
\maketitle

\begin{abstract}
Weak gravitational lensing convergence peaks, the local maxima in weak lensing convergence maps, have been shown to contain valuable cosmological information complementary to commonly used two-point statistics. To exploit the full power of weak lensing for cosmology, we must model baryonic feedback processes because these reshape the matter distribution on non-linear and mildly non-linear scales. We study the impact of baryonic physics on the number density of weak lensing peaks using the FLAMINGO cosmological hydrodynamical simulation suite. We generate ray-traced full-sky convergence maps mimicking the characteristics of a Stage IV weak lensing survey. We compare the number densities of peaks in simulations that have been calibrated to reproduce the observed galaxy mass function and cluster gas fraction or to match a shifted version of these, and that use either thermally driven or jet AGN feedback. We show that the differences induced by realistic baryonic feedback prescriptions (typically $5-30$\% for $\kappa = 0.1-0.4$) are smaller than those induced by reasonable variations in cosmological parameters ($20-60$\% for $\kappa = 0.1-0.4$) but must be modeled carefully to obtain unbiased results. The reasons behind these differences can be understood by considering the impact of feedback on halo masses, or by considering the impact of different cosmological parameters on the halo mass function. Our analysis demonstrates that, for the range of models we investigated, the baryonic suppression is insensitive to changes in cosmology up to $\kappa \approx 0.4$ and that the higher $\kappa$ regime is dominated by Poisson noise and cosmic variance.
\end{abstract}

\begin{keywords}
gravitational lensing: weak -- cosmology: theory -- methods: numerical -- large-scale structure of the Universe
\end{keywords}



\section{Introduction}
Over the last few decades, the spatially flat $\Lambda$CDM model has become the generally accepted standard cosmological model. While it depends on only six free parameters, it can predict several key observations with great accuracy, including fluctuations of the cosmic microwave background (CMB) \citep[][]{Planck2020}{}{}, galaxy clustering \citep[][]{Anderson2014}{}{}, and type Ia supernovae \citep[][]{Abbott2019}{}{} \citep[for a recent review see][]{Lavah2022}{}{}. As the observations become increasingly more constraining, tensions between different cosmological probes have emerged, most notably on the value of the Hubble constant, $H_\mathrm{0}$, and $S_\mathrm{8}$ \citep[e.g.][]{Hildebrandt2017,Abdalla2022,Schoneberg2022,Clark2023}{}{}. Understanding the origin of these tensions, which may lead to new physics beyond the $\Lambda$CDM model, is one of the major goals of modern cosmology.

One of the key tools to constrain cosmology is cosmic shear, the slight distortion of distant galaxy images through weak gravitational lensing (weak lensing or WL) by the large-scale structure (LSS) of the Universe. It allows us to estimate the projected matter distribution along the line of sight, which can be related to the underlying cosmology. For reviews, see \citet{bartelmann2001weak,hoekstra2008weak,kilbinger2015cosmology}. Upcoming next-generation weak gravitational lensing surveys, carried out from space by the Roman \citep[][]{spergel2015wide} and the recently-launched \textit{Euclid} \citep[][]{laureijs2011euclid} satellites, and from the ground by Rubin \citep[][]{abell2009lsst}, collectively referred to as Stage IV surveys, will cover almost the entirety of the observable sky suitable for WL. They will reach unprecedented depths as well as measure the WL signal as a function of redshift, allowing them to quantify the evolution of the matter distribution in the Universe. Jointly, these missions aim to provide insight into the nature of dark matter, dark energy, and the expansion history of the Universe.

Typically, WL surveys use two-point statistics, either in configuration space or harmonic space, to constrain the cosmological model \citep[e.g.][]{asgari2021kids,amon2022dark,hamana2020cosmological}{}{}. Astrophysical feedback processes (e.g. supernova explosions, and active galactic nuclei (AGN)) reshape the matter distribution on partly the same scales that are typically probed by the two-point inferences, thereby complicating the analysis. It has been shown that this so-called baryonic feedback suppresses the matter power spectrum on scales of $k \lessapprox 0.1~h~$Mpc$^{-1}$ and enhances the power on even smaller scales \citep[e.g.][]{vDaalen2011baryons,vDaalen2020baryons,Chisari2018,Schneider2020,schaye2023flamingo,Salcido2023}{}{}. When not taking these baryonic effects into account, strong biases in inferences may arise for WL statistics \citep[e.g.][]{Semboloni2011,Semboloni2013,Gouin2019Horizon,weiss2019peaks}{}{}.

Two-point statistics encapsulate all cosmological information in the underlying field if it can be described as a Gaussian random field (GRF). While today this applies to the largest scales, the cosmological information on smaller scales, which correspond to the regime of non-linear collapse and contain additional information, is not fully captured by Gaussian statistics. Over the last decade, interest has grown in non-Gaussian statistics, which are able to probe this regime. Examples of commonly used beyond-Gaussian statistics are the bispectrum \citep[e.g.][]{dodelson2005weak}{}{}, Minkowski functionals \citep[e.g.][]{kratochvil2012probing}{}{}, higher order moments of the convergence field \citep[e.g.][]{petri2013cosmology}{}{}, WL peaks or voids \citep[e.g.][]{kratochvil2010probing,davies2021constraining,davies2022cosmological}{}{}, and Betti numbers \citep[e.g.][]{feldbrugge2019stochastic}{}{}. The addition of non-Gaussian statistics can provide tighter cosmological constraints \citep[e.g.][]{Euclid2023HOWLS}{}{} and help discriminate between cosmological and baryonic effects \citep[e.g.][]{Semboloni2013}{}{}. In general, the baryonic impact on these statistics is not understood as well as it is for the two-point statistics and only approximate methods, for example by using a halo model \citep[e.g.][]{Sabyr2022,Asgari2023}{}{}, for calculating these quantities exist. 

In this paper, we choose to focus on one of these non-Gaussian statistics, namely WL peaks, which correspond to local maxima in the WL convergence field. WL peaks have been found to be highly complementary to typical 2-point statistics, both in simulations \citep[e.g.][]{dietrich2010cosmology}{}{} and observations \citep[e.g.][]{Marques2023peaks}{}{}. Whereas peaks can arise due to chance alignments of haloes along the line of sight, the highest peaks primarily stem from a single high-mass halo along the line of sight \citep[][]{li2019constraining}{}{}. Lower peaks are typically caused by multiple smaller haloes aligned along the line of sight, but dominate the cosmological information contained in the peaks \citep[][]{yang2011cosmological}{}{}. Peak counts are thus directly sensitive to the number of haloes and therefore directly probe the halo mass function (HMF), which depends strongly on cosmology \citep[e.g.][]{kaiser1986evolution,tinker2008toward,mcclintock2019aemulus}{}{}, such that changes in cosmology directly influence the number density of WL peaks. Compared to other probes of high-density regions, such as cluster X-ray luminosity and temperature, or cluster optical richness, WL peak counts are a more direct tracer of the total mass in haloes, and they are not plagued by uncertainties arising from a set of assumptions regarding the dynamical state of the galaxy clusters (e.g. hydrostatic equilibrium or relaxedness), nor do they require scaling relations between mass and luminosity tracers. Therefore, peaks are an ideal probe to enhance WL inferences in constraining cosmology.

There are two main approaches to studying the impact of baryons on weak lensing statistics. The first approach, which we adopt, uses full-hydrodynamical simulations employing sub-grid models of relevant baryonic processes. Despite being computationally more expensive, this method has the important advantage of being fully self-consistent and allowing to compare the WL signal to non-gravitational probes such as X-ray and SZ. The second is to use $N$-body simulations to model the evolution of dark matter, and to modify the shape of these dark matter only (DMO) simulations using a baryonic correction model (BCM) \citep[e.g.][]{schneider2019quantifying,arico2020modelling}{}{}.  \citet{Lee2023BCMtohydro} compared an $N$-body + BCM and its corresponding hydrodynamical simulation and found that current BCM approaches that are calibrated on the power spectrum are not flexible enough for a WL peak count inference for Stage IV WL surveys, which is the focus of this paper. However, it is not a priori expected that a calibration on the power spectrum can recover all the peak properties as not all information in the peaks is captured by the power spectrum. \citet{lu2021impact} use a similar but adapted BCM in the context of a Stage III inference and they found that the degeneracy between cosmological and baryonic parameters may be broken by considering peaks combined with the power spectrum. 

Peak counts have been studied in the context of hydrodynamical simulations before. In general, it has been found that peak counts are sensitive to the baryonic contribution, and hence, for cosmological inferences based on peak counts, baryons have to be considered \citep[e.g.][]{yang2013baryon,coulton2020weak}{}{}. Similarly, the impact of neutrinos on peak counts has been studied using the BAHAMAS simulations \citep[][]{mcchartybahamas2017,mccarthy2018bahamas}{}{}. \citet{fong2019impact} found that depending on the neutrino mass, either baryonic or neutrino effects dominate, and both effects should be accounted for in a proper WL peak analysis. Recently, \citet{FerlitoMTNGconvergence} carried out a comparison of WL peaks in cosmological hydrodynamical simulations, primarily focussing on WL convergence maps constructed using the MilleniumTNG (MTNG) simulation suite \citep[][]{Pakmor2022MTNG}{}{}. Their analysis focuses on the contribution of neutrinos to WL peak counts. While we also look at the impact of massive neutrinos, we concentrate on the impact of baryonic feedback processes on WL peaks. \citet{FerlitoMTNGconvergence} also studied the baryonic impact on WL peaks by comparing their simulations to their corresponding DMO runs, as well as with simulations from the literature. As these simulations may differ significantly in terms of code, subgrid physics, cosmology, and resolution, a direct interpretation of the impact of the baryonic effect is not straightforward. Here, we instead use a consistent suite of simulations that systematically varies baryonic feedback strength. 

In this paper, we explore the impact of baryonic physics on WL peak counts in the FLAMINGO simulation suite \citep[][]{schaye2023flamingo,kugel2023flamingo}{}{}. The hydrodynamical simulations were calibrated, using machine-learning techniques, to reproduce the observed present-day gas fractions in clusters and the galaxy stellar mass function. The suite includes separately calibrated models that systematically vary these observables. In this way, we can directly relate changes in key observables, induced by feedback variations, to differences in WL peak counts. Additionally, the suite contains variations in cosmological parameters and neutrino masses, allowing us to compare the impact of cosmology to that of baryonic physics. To be able to quantify the changes induced by astrophysical feedback processes on next-generation WL surveys, we carry out a dedicated full-sky analysis at a high angular resolution in which the characteristics of a Stage IV WL survey are incorporated. The signal is determined for virtual observers which were placed within the simulation volume using a backward ray-tracing methodology and spherical harmonics on the full-sky sphere. 

This paper is organized as follows. In Section~\ref{ch:theory} we introduce the relevant WL and ray-tracing theory. Section~\ref{ch:methods} introduces the FLAMINGO simulation suite, the key features of the different baryonic and cosmology models, and the construction of the WL maps. Here, we also validate our approach by quantifying the sensitivity to different choices of smoothing, interpolation scheme, and angular resolution. In Section~\ref{ch:results} we start by quantifying our level of numerical convergence in terms of box size and numerical resolution and the impact of cosmic variance on our signal, finding that the measured peak statistics are robust. Then, we compare simulation variations that were calibrated to different values of gas fractions in clusters and the galaxy stellar mass function. In this way, we are able to understand the relation between stronger feedback, leading to lower gas fractions in clusters or lower galaxy stellar mass functions, and the observed WL peak distribution. We then compare the different AGN subgrid models and the different cosmology variations in Section~\ref{ch:discussion}, where we also look at the separability of the baryonic and cosmological effects. We conclude by comparing the differences induced by baryonic feedback variations to those due to changes in cosmology. Our main results are summarized in Section~\ref{ch:conclusion}.

\section{Weak Lensing Theory}\label{ch:theory}
We will generate WL convergence signals from pixelized surface mass density maps discretized in redshift as seen by a virtual observer within a simulation. Before describing the construction of the convergence maps, we first summarize the main equations relevant to WL. For an extensive recent review, see \citet{kilbinger2015cosmology}.
We assume a flat FLRW metric such that the comoving angular diameter distance, $f_k(\chi)$, equals the comoving line of sight distance, $\chi$. In WL, where deflections are small, the deflection field, $\boldsymbol{\alpha}$, which at each angular position $\boldsymbol{\theta}$ describes the change in the position of a light ray perpendicular to the direction of travel, can be expressed in terms of the Newtonian gravitational potential $\Phi$ as 
\begin{equation}
    \mathrm{d}  \boldsymbol{\alpha} = -\frac{2}{c^2} \nabla_{\perp} \Phi \,   \mathrm{d}\chi,
\end{equation}
where $c$ is the speed of light and the gradient, $\nabla_\perp =$ ($\partial/\partial\beta_1,\partial/\partial\beta_2$), is evaluated perpendicular to the light ray's direction of propagation. Integration over comoving distance yields the angular position of the light rays as seen by the observer,
\begin{equation}\label{eqn:deflec_int}
   \boldsymbol{\beta}(\boldsymbol{\theta},\chi) = \boldsymbol{\beta}(\boldsymbol{\theta},0) -\frac{2}{c^2} \int_0^\chi   \mathrm{d}\chi'\, \frac{\chi-\chi'}{\chi\chi'} \, \nabla_\perp \Phi(\boldsymbol{\beta}(\boldsymbol{\theta},\chi'),\chi'),
\end{equation}
where $\boldsymbol{\beta}(\boldsymbol{\theta},0) = \boldsymbol{\theta}$ is the observed angular position. The deflection of the photon path as it passes along a non-uniform matter distribution is described by the distortion matrix $A$, which is the derivative of the angular position and given by
\begin{equation}\label{eqn:magmat_int}
    A_{ij}(\boldsymbol{\theta},\chi) \equiv \frac{\partial\boldsymbol{\beta}}{\partial \boldsymbol{\theta}} = \delta_{ij} -\frac{2}{c^2} \int_0^\chi \mathrm{d}\chi'\, \frac{\chi-\chi'}{\chi\chi'}\, \partial_i \partial_j \Phi(\boldsymbol{\beta}(\boldsymbol{\theta},\chi'),\chi'),
\end{equation}
where $i$ and $j$ are taken over the two angular coordinates on the sphere and $\delta_{ij}$ is the Kronecker delta function. Conventionally, the matrix is decomposed as 
\begin{align}\label{eqn:magmat_matrix}
\begin{split}
    A &\equiv \begin{pmatrix}
\cos \omega & \sin \omega \\
-\sin \omega  & \cos \omega 
\end{pmatrix} \begin{pmatrix}
1-\kappa - \gamma_1 & -\gamma_2 \\
-\gamma_2  & 1 - \kappa + \gamma_1 
\end{pmatrix} \\ 
&\approx \begin{pmatrix}
1-\kappa - \gamma_1 & -\gamma_2 + \omega \\
-\gamma_2 -\omega & 1 - \kappa + \gamma_1 
\end{pmatrix},
\end{split}
\end{align}
which we refer to as the magnification matrix. Here, we assumed that the image rotation, described by the rotation angle $\omega$, is small, as shown by \citet{jain2000ray}. $\kappa$ is the lensing convergence, which we will use to quantify the WL strength and $\gamma = \gamma_1 + \mathrm{i} \gamma_2$ is the lensing shear.

In WL, where the deflection angles are small, a common approach is to evaluate the deflection field on unperturbed photon paths, in which case the dependence of the angular position on comoving distance vanishes (i.e. $\boldsymbol{\beta}(\boldsymbol{\theta},\chi) \cong \boldsymbol{\beta}(\boldsymbol{\theta})$). This approximation is referred to as the Born approximation. The accuracy of the Born approximation has been well established for the reconstruction of the convergence angular power spectrum and cosmological parameter inferences based on it \citep[e.g.][]{giocoli2016multidarklens,hilbert2020accuracy}{}{}. For non-Gaussian statistics, the impact and validity of the Born approximation remain uncertain, as it has been found to impact galaxy–galaxy lensing shear profiles \citep[][]{simon2018scale}{}{}, the CMB lensing bispectrum \citep[][]{pratten2016impact}{}{}, and higher order moments of convergence profiles, where \citet{petri2017validity} find a $2.5\sigma$ bias on cosmological parameter estimations for a Rubin-like survey. Similarly, \citet{lu2021impact} find that parameter inferences based on the Born approximation and peak counts can be up to $2\sigma$ biased for Hyper Suprime-Cam-like surveys. 

As our analysis aims to mimic a \textit{Euclid}-like signal, which will give tighter constraints, we expect this bias to be even more significant, stressing the need to adopt a beyond-Born approach. In our analysis, we will use the FLAMINGO mass shells. These shells are discrete full-sky maps spaced regularly in redshift ($\Delta z = 0.05 $ between $z=0$ and $3$). To analyze the maps, the theoretical equations \ref{eqn:deflec_int} and 
\ref{eqn:magmat_int} need to be discretized. We choose to apply a backward ray-tracing method, as introduced by \citet{jain2000ray}. Here, starting from the position of a virtual observer, we track a ray's trajectory back in time as it is deflected by the gravitational potential of discrete matter shells along its propagation. For a detailed description of the deflection of the rays using discrete shells, see \citet{becker2013calclens}. 

In summary, the expressions for the deflection angle and magnification matrix (eqs.~\ref{eqn:deflec_int} and~\ref{eqn:magmat_int}) can be expressed as a sum over a discrete number of lensing planes such that at the $n$-th plane they are given by
\begin{align} \label{eqn:deflec_field1}
    \boldsymbol{\beta}^n = \boldsymbol{\beta}^0 - \sum_{m=0}^{n-1} \frac{\chi^{n}-\chi^m}{\chi^n}\, \boldsymbol{\alpha}^m,
\end{align} 
and
\begin{align} \label{eqn:mag_mat1}
   A_{ij}^n = \delta_{ij} - \sum_{m=0}^{n-1} \frac{\chi^{n}-\chi^m}{\chi^n}\, U_{ik}^{m}\, A_{kj}^{m},
\end{align} 
where $\chi^m$ is the comoving distance to the $m$-th plane. For clarity, we left out the dependence of the quantities on $\chi$ and $\boldsymbol{\theta}$. Here, we defined the shear matrix $U$ such that the deflection field ($\boldsymbol{\alpha}^m$) and shear matrix equal the first- ($  \boldsymbol{\alpha}^n(\boldsymbol{\theta}) = \nabla_{\perp} \psi^n(\boldsymbol{\theta})$) and second-order ($U_{ij}^n = \partial_i \partial_j \psi^n$) derivatives, orthogonal to the direction of propagation, of the lensing potential $\psi$, which is defined as 
\begin{align}\label{eqn:lensing_potential}
\psi^n \equiv \frac{2}{\chi^n c^2} \int_{\chi_{\mathrm{min}}}^{\chi_{\mathrm{max}}} \mathrm{d}\chi\, \Phi,
\end{align}
where $\chi_{\mathrm{min}}$ and $\chi_{\mathrm{max}}$ are the comoving distances to the beginning and end of the shell, respectively. In principle, using Equations~\ref{eqn:deflec_field1} and \ref{eqn:mag_mat1}, the magnification matrix and angular positions at the next plane are evaluated using those of all previous planes. It is, however, computationally infeasible to construct full-sky maps with a resolution sufficiently high for the analysis of Stage IV WL surveys in this way. \citet{hilbert2009ray} showed that the sum can be combined into a recurrence relation where only the information of the previous two shells needs to be stored in memory. The recurrence relation is a result of the exact form of the transverse comoving distance in a generic Roberston-Walker metric and is given by \citep[][]{schneider2016generalized}{}{}:
\begin{align}\label{eqn:mag_mat_recurr}
\begin{split}
    A^{n+1}_{ij} &= \bigg(1-\frac{\chi^n}{\chi^{n+1}}\frac{\chi^{n+1}-\chi^{n-1}}{\chi^{n}-\chi^{n-1}}\bigg)\, A^{n-1}_{ij} \\ &+\frac{\chi^n}{\chi^{n+1}}\frac{\chi^{n+1}-\chi^{n-1}}{\chi^{n}-\chi^{n-1}} \, A^{n}_{ij}- \frac{\chi^{n+1}-\chi^{n}}{\chi^{n+1}} \, U^{n}_{ik}\, A^{n}_{kj},
\end{split}
\end{align}
with the initialisation given by $A^{-1}_{ij}$ = $A^{0}_{ij}$ = $\delta_{ij}$. A similar recurrence relation holds for the angular position, where the positions at the next plane ($\boldsymbol{\beta}^{n+1}$) can be evaluated using the position at the current plane ($\boldsymbol{\beta}^{n}$) and the previous plane ($\boldsymbol{\beta}^{n-1}$),
\begin{align}\label{eqn:ang_pos_recur}
\begin{split}
    \boldsymbol{\beta}^{n+1} &= \bigg(1-\frac{\chi^n}{\chi^{n+1}}\frac{\chi^{n+1}-\chi^{n-1}}{\chi^{n}-\chi^{n-1}}\bigg)\, \boldsymbol{\beta}^{n-1} \\ &+ \frac{\chi^n}{\chi^{n+1}}\frac{\chi^{n+1}-\chi^{n-1}}{\chi^{n}-\chi^{n-1}} \, \boldsymbol{\beta}^{n} - \frac{\chi^{n+1}-\chi^{n}}{\chi^{n+1}}\, \boldsymbol{\alpha}^{n},
\end{split}
\end{align}
where $\boldsymbol{\beta}^{-1} = \boldsymbol{\beta}^{0} = \boldsymbol{\theta}$ are the initial angular positions of the rays. In Section~\ref{sec:conv_maps} we will explain how the FLAMINGO lightcones can be related to the deflection field and shear matrix in Equations~\ref{eqn:mag_mat_recurr}~\&~\ref{eqn:ang_pos_recur}. The equations can then be combined with Equation~\ref{eqn:magmat_matrix} to estimate the WL convergence, $\kappa$.

\section{Methods}\label{ch:methods}
\subsection{The FLAMINGO simulations}
For our analysis, we make use of the FLAMINGO simulation suite, a recent collection of large cosmological hydrodynamical simulations explicitly designed for the purpose of large-scale structure analysis and cluster physics. For a full description of the simulation details, its performance with respect to observables, and calibration strategy see \citet{schaye2023flamingo}, \citet{kugel2023flamingo} and \citet{McCharty2023}. We summarize here the key elements.

The simulations were run using the SWIFT hydrodynamics code \citep[][]{schaller2023swift}{}{} with the SPHENIX smoothed particle hydrodynamics (SPH) implementation \citep[][]{borrow2022sphenix}{}{}. Neutrinos are modeled as massive particles, using the $\delta f$ method of \citet{elbers2021optimal} that was designed to reduce particle shot noise. The simulations include radiative cooling and heating that is implemented on an element-by-element basis \citep[][]{Ploeckinger2020rates}{}{}, star formation \citep[][]{Schaye2008stars}{}{}, and time-dependent stellar mass loss as described by \citet{Wiersma2009sml}. Supernova (SN) and stellar feedback are implemented kinetically \citep[][]{DallaVecchia2008SNfeedback}{}{} by kicking neighboring particles in a way that conserves energy as well as linear and angular momentum, as described by 
\citet{Chaikin2023SNfeedback}. The accretion of gas onto supermassive black holes and the subsequent thermal AGN feedback is described in \citet{Booth2009AGN} and the kinetic jet feedback is based on the AGN jet implementation of \citet{Husko2022jets}, where gas particles receive a kick to a fixed target jet velocity in the direction given by the spin of the BH. 

An important novel feature in the suite is that the runs have been calibrated using Gaussian process emulators trained on Latin hypercubes of smaller simulations where subgrid parameters are varied \citep[][]{kugel2023flamingo}{}{}. In this way, to study the impact of feedback variations, instead of varying a single, unobservable parameter relating to the specific implementation of a feedback process, a set of subgrid parameters is systematically varied by fitting a Gaussian process emulator such that the simulations can be characterized by (shifts in) real observables instead of subgrid parameters. The FLAMINGO variations were calibrated to the observed $z=0$ galaxy stellar mass function (SMF) and gas fractions in low$-z$ clusters. Comparing sets of simulations calibrated to different observables allows for more instructive comparisons than comparing simulations that differ in specific subgrid parameters. Additionally, expected observational biases were included in the calibration. For the study of WL peaks, which trace the total mass in objects, calibrating to these observables ensures that the objects causing the peaks have a realistic ratio of gas and stars to DM. In the suite, the four subgrid parameters that are varied relate to the subgrid prescription of the supernova and AGN feedback, namely the fraction of stellar energy feedback that couples to the ISM ($f_{\mathrm{SN}}$), the target wind velocity for supernova feedback ($\Delta v_{\mathrm{SN}}$), the black hole (BH) accretion rate boost factor ($\beta_{\mathrm{BH}}$) and either the AGN heating temperature ($\Delta T_{\mathrm{AGN}}$) or jet velocity ($v_{\mathrm{jet}}$) for thermal and kinetic jet AGN feedback, respectively. 

At fixed fiducial feedback calibration, the suite has 4 different runs varying the mass resolution and box size. The details of these runs are listed in Table~\ref{tab:vars}. The identifier of each run indicates the box size in comoving Gpc (cGpc) and the rounded $\log_{\mathrm{10}}$ mass of the baryonic particle mass. For example, the flagship run (L2p8$\_$m9) is a 2.8\,cGpc box with 0.3 trillion ($5040^3$ for dark matter (DM) and baryons and $2800^3$ for massive neutrinos) particles with a baryonic particle mass of $1.07~\times~10^9\,\mathrm{M_\odot}$, making it the cosmological hydrodynamical simulation with the highest number of resolution elements run to $z=0$ to date. For each of the models, a corresponding DMO$+\nu$ run, with the same initial phases, exists, whose identifier carries the postfix `$\_$DMO'. The cosmology of these runs was taken from the Dark Energy Survey year three `3$\times$2pt+ All Ext.' $\Lambda$CDM cosmology \citep[][]{abbott2022dark}{}{}, indicated as `D3A' in Table~\ref{tab:cosmologies}. The initial conditions were generated using \textsc{MonofonIC} \citep[][]{hahn2020monofonic,michaux2021monofonic}{}{} using 3-fluid third-order Lagrangian perturbation theory with separate transfer functions for baryons, cold dark matter (CDM), and neutrinos \citep[][]{Elbers2022}{}{}. The simulations are initiated at a redshift of $z=31$.

\begin{table}
\centering
\caption{Characteristics of the 4 simulation variations that differ only in numerical resolution and box size. The columns indicate the simulation identifier; the box size length in comoving Gpc, $L$; the number of baryonic and DM particles, $N_\mathrm{b}$; the number of massive neutrino particles, $N_\nu$; the initial mean baryonic particle mass, $m_\mathrm{b}$; and the mean CDM particle mass, $m_{\mathrm{CDM}}$.}
\label{tab:vars}
\begin{tabular}{lllllll}
\hline
Identifier & $L$    & $N_\mathrm{b}$  & $N_\nu$  & $m_\mathrm{b}$               & $m_{\mathrm{CDM}}$  \\
           & cGpc &          &          & M$_\odot$        & M$_\odot$                           &                   \\
           \hline
L1$\_$m8   & 1      & 3600$^3$ & 2000$^3$ & $1.38\times 10^{8}$ & $7.06\times 10^{8}$                                    \\
L1$\_$m9   & 1      & 1800$^3$ & 1000$^3$ & $1.07\times 10^{9}$ & $5.65\times 10^{9}$                                    \\
L1$\_$m10  & 1      & 900$^3$  & 500$^3$  & $8.56\times 10^{9}$ & $4.52\times 10^{10}$                                 \\
L2p8$\_$m9 & 2.8    & 5040$^3$ & 2800$^3$ & $1.07\times 10^{8}$ & $5.65\times 10^{8}$                                    \\
\hline
\end{tabular}
\end{table}

\subsection{Model variations}\label{sec:baryonic_variations}
The suite includes 12 model variations in L1 boxes. The details of all the runs varying the baryonic feedback model with respect to the fiducial 1\,cGpc box are listed in Table~\ref{tab:bar_vars}. At fixed cosmology, 8 runs were calibrated to different galaxy stellar mass functions (M*) and/or gas fractions in clusters ($f_{\mathrm{gas}}$) and/or differ in overall AGN subgrid feedback prescription. The subgrid parameter values have been chosen such that the stellar mass function and/or gas fraction in clusters within the simulation are a set number of standard deviations from the fiducial model ($\Delta$M* and $\Delta f_{\mathrm{gas}}$, respectively), as indicated in the second and third columns of Table~\ref{tab:bar_vars}. The standard deviation on the gas fractions was estimated by bootstrapping the X-ray data \citep[table~5 of][]{kugel2023flamingo}{}{} and the error on the weak lensing data \citep[][]{Akino2022}{}{}. Similarly, the shift of the SMF is the expected systematic error (0.14~dex) on the stellar masses from \citet{Behroozi2019}. The suite includes more models with stronger than weaker feedback to allow quantifying the observational signatures of exceptionally strong feedback, which affects larger length scales than exceptionally weak baryonic feedback.

There are 4 runs varying only the gas fractions in clusters, whose identifiers are given by `fgas$\pm$n$\sigma$', where n is the number of standard deviations by which the gas fraction is varied. One run was calibrated to match the galaxy stellar mass function shifted to a lower mass by $1\sigma$ (M*$-\sigma$). The M*$-\sigma\_$fgas$-4\sigma$ run varies both observables. The fiducial implementation of AGN feedback is thermal, but there are two models with a kinetic jet AGN feedback description, which are denoted by `Jet' and `Jet$\_$fgas$-4\sigma$', where for the latter the target gas fraction in clusters is reduced. The Jet models can be used to access the sensitivity of observables to variations in subgrid models calibrated to the same observables. 

At fixed baryonic calibration, there are 4 cosmology variations. The different cosmologies and their cosmological parameters are listed in Table~\ref{tab:cosmologies}. In addition to the fiducial D3A cosmology that was used for all the different baryonic physics runs, there are 3 variations based on cosmic microwave background (CMB) measurement from Planck. The first CMB cosmology is the best-fit \citet{Planck2020} $\Lambda$CDM cosmology with $\sum m_\nu = 0.06$ eV (Planck). The other two Planck cosmology variations include heavier neutrinos (3 species each with $m_\nu c^2$ = 0.08 eV), one in which the other parameters are changed according to their best-fit values within the Planck MCMC chains (PlanckNu0p24Var), and one in which the other parameter values are fixed and only $\Omega_{\mathrm{CDM}}$ is adjusted to keep $\Omega_{\mathrm{m}}$ fixed (PlanckNu0p24Fix). The final variation is a cosmology model with a lower value of $S_8$ (LS8), taken from \citet{amon2023consistent}.

Critically, all variations in the L1 box have been run using the same initial phases. This allows us to isolate the effect of the varying baryonic physics and cosmology from cosmic variance induced by different initial realizations.

\begin{table}
\centering
\caption{The baryonic physics variations in the 1\,cGpc FLAMINGO box. For each of the different models, indicated by their identifier, the second and third columns indicate the number of observational standard deviations ($\sigma$) by which the galaxy stellar mass function (M*) and gas fractions in clusters ($f_{\mathrm{gas}}$) in the simulation are shifted compared to the fiducial L1$\_$m9 model, respectively. The final column indicates the method of AGN feedback in the run. All the runs with different baryonic feedback were run using the same D3A cosmology (see Table~\ref{tab:cosmologies}) and with the same initial conditions.}
\label{tab:bar_vars}
\begin{tabular}{lllll}
\hline
Identifier                      & $\Delta$M*  & $\Delta f_{\mathrm{gas}}$ & AGN mode        \\
                                &   $\sigma$ &   $\sigma$ & \\
\hline
L1$\_$m9                        & 0                                                                                                    & 0                                                                                      & thermal             \\
fgas$+2\sigma$                  & 0                                                                                                    & +2                                                                                     & thermal              \\
fgas$-2\sigma$                  & 0                                                                                                    & -2                                                                                     & thermal             \\
fgas$-4\sigma$                  & 0                                                                                                    & -4                                                                                     & thermal              \\
fgas$-8\sigma$                  & 0                                                                                                    & -8                                                                                     & thermal             \\
M*$-\sigma$                 & -1                                                                                                   & 0                                                                                      & thermal              \\
M*$-\sigma\_$fgas$-4\sigma$ & -1                                                                                                   & -4                                                                                     & thermal              \\
Jet                             & 0                                                                                                    & 0                                                                                      & jet               \\
Jet$\_$fgas$-4\sigma$           & 0                                                                                                    & -4                                                                                     & jet               \\

\hline
\end{tabular}
\end{table}

\begin{table*}
\caption{The cosmological parameter values used in the different FLAMINGO simulations. The columns indicate the cosmology identifier; the dimensionless Hubble constant, $h = H_0 / (100$ km/s/Mpc$)$; the total matter density parameter, $\Omega_\mathrm{m}$; the baryonic matter density parameter, $\Omega_\mathrm{b}$;
the cosmological constant density parameter, $\Omega_\Lambda$; the neutrino mater density parameter, $\Omega_\nu$; the summed mass of the massive neutrino species, $\sum m_\nu c^2$; the amplitude of the primordial power spectrum, $A_\mathrm{s}$; the power-law index of the primordial matter power spectrum, $n_\mathrm{s}$, the amplitude of the linear theory power spectrum parameterized as the r.m.s. mass density fluctuation in spheres of radius 8 $h^{-1}$~Mpc at $z=0$, $\sigma_8$, and the amplitude of the initial power spectrum parametrized as $S_8 \equiv \sigma_8 \sqrt{\Omega_\mathrm{m}/0.3}$.}
\label{tab:cosmologies}
\begin{tabular}{lllllllllll}
\hline
Cosmology       & $h$   & $\Omega_\mathrm{m}$ & $\Omega_\mathrm{b}$ & $\Omega_\Lambda$ & $\Omega_\nu$      &  $\sum m_\nu c^2$ & $A_\mathrm{s}$            & $n_\mathrm{s}$ & $\sigma_8$ & $S_8$ \\
                &       &            &                  &            & $\times 10^{-3}$ & [eV]               & $\times 10^{-9}$ &       &            &       \\
\hline
D3A             & 0.681 & 0.306      &  0.0486           &0.694     & 1.39             & 0.06             & 2.099            & 0.967 & 0.807      & 0.815 \\
Planck          & 0.673 & 0.316      &   0.0494          & 0.684     & 1.42             & 0.06             & 2.101            & 0.966 & 0.812      & 0.833 \\
PlanckNu0p24Var & 0.662 & 0.328      &   0.0510         & 0.672      & 5.87             & 0.24             & 2.109            & 0.968 & 0.772      & 0.807 \\
PlanckNu0p24Fix & 0.673 & 0.316      &   0.0494         & 0.684      & 5.69             & 0.24             & 2.101            & 0.966 & 0.769      & 0.789 \\
LS8             & 0.682 & 0.305      &  0.0473          & 0.695     & 1.39             & 0.06             & 1.836            & 0.965 & 0.760      & 0.766\\
\hline
\end{tabular}
\end{table*}

\subsection{Lightcones}
To construct the WL convergence maps, we use the mass maps of the FLAMINGO lightcones. The lightcones correspond to virtual observers within the FLAMINGO simulation suite. They have been constructed by recording particles crossing the past lightcone of a virtual observer. The lightcones of the L1 runs are provided as 60 projected shells spaced equally in redshift between $z = 0$ and $3$ (i.e. $\Delta z = 0.05$). Particles within the innermost 3~Mpc have been removed. The shells are stored as HEALPix maps \citep[][]{gorski2005healpix}{}{} with a resolution of $N_{\mathrm{side}}$ = 16384 (with a number of pixels of $N_\mathrm{pix} = 12N_{\mathrm{side}}^2$), corresponding to an angular resolution of 0.21 arcmin. For the details of the lightcone construction and properties, see Appendix~A of \citet{schaye2023flamingo}.

As the box size is not large enough to cover the distance up to $z=3$, which is the range we consider (see Section~\ref{sec:nz_source}), the box is replicated to reach this distance. For the L2p8 and L1 boxes, this requires 5 and 12 additional replications, respectively. In Appendix~\ref{app:box_rotation}, we compare different ways of implementing the box replication, including different mass shell rotation strategies. We compare the signal in the L1$\_$m10$\_$DMO run using non-rotated shells, the case in which every shell is randomly rotated and correlations along the line of sight may thus be unnecessarily erased, and the case of rotating the shells whenever the lightcone diameter is larger than the box length. We compare the measurements to those for a 5.6\,cGpc DMO run (L5p6$\_$m10$\_$DMO), which can cover the lightcone diameter without replications up to $z=0.8$. We find only minor differences between the different box rotation strategies. To prevent encountering the same structure multiple times, but also to not unnecessarily erase correlations along the line of sight, we choose to randomly rotate the shells every half box length. The same random angles are applied to all L1 observers. As these observers are all placed at the same position and reside in boxes with the same initial phases, we can directly study the impact of both the cosmology variations and baryonic physics on the measured WL signal generated by the same objects. Within the L2p8$\_$m9 box, 8 observers have been placed at the coordinates ($\pm L$/4,$\pm L$/4,$\pm L$/4), where $L$ is the simulation box size. These lightcones have 68 shells up to $z=5$ but we only use the first 60 lightcones to facilitate a direct comparison with the L1 lightcones, as the first 60 shells have the same redshift spacing as the lightcones of the L1 runs. The mean shell thickness is $\approx110$~Mpc. \citet{Zorrilla2020} found that the bias resulting from such a redshift discreteness is statistically insignificant in the presence of shape noise for Stage IV cosmic shear statistics. We compare the 8 different lightcones to quantify the impact of cosmic variance on our analysis. Additionally, we carry out numerical convergence tests for simulation box size and resolution.

\subsection{Full-sky WL convergence maps}\label{sec:conv_maps}
\subsubsection{Spherical harmonics}\label{sec:spherical_harmonics}
To exploit the full power of the all-sky FLAMINGO HEALPix maps, we carry out our analysis in spherical harmonics space. Because some Fortran compilers do not support 64-bit array sizes, the HEALPix library officially does not support maps with a size larger than $N_{\mathrm{side}}=8192$. Even though our analysis uses the Python implementation of the HEALPix library \citep[][]{Zonca2019Healpy}{}{}, which is technically not affected by this issue, some internal functionality that we need relies on 32-bit indexing, which means we cannot use these functions on the full-resolution FLAMINGO lightcone maps. We therefore limit our analysis to downsampled maps with $N_{\mathrm{side}}$ = 8192, corresponding to a pixel size of 0.43 arcmin for a full-sky analysis. As we smooth our final maps with a Gaussian kernel with a full width at half maximum  (FWHM) of 1 arcmin, we are not directly limited by the resolution of the HEALPix maps, which we illustrate in more detail in Section~\ref{sec:validation}. At each of the lightcone shells, we determine the lensing potential (given by Equation~\ref{eqn:lensing_potential}) using the solution of the two-dimensional Poisson equation which we define as the convergence at plane $n$ ($K^n$) as 
\begin{align} \label{eqn:conv_at_shell} 
   K^n(\boldsymbol{\theta}) \equiv \nabla^2 \psi^n = \frac{3H_0^2\Omega_{\mathrm{m}}}{2 c^2}\, \chi^n \, (1+z^n) \, \Delta \chi^n \, \delta^n(\boldsymbol{\theta}),
\end{align}
where $\Delta \chi = \chi_{\mathrm{max}} -  \chi_{\mathrm{min}} $ is the thickness of the shell and we choose to evaluate $
\chi^n$ and $z^n$ at the comoving center of each shell. The overdensity $\delta^n(\boldsymbol{\theta})$ can be directly evaluated from the surface mass density and is given by 
\begin{align}
    \delta^n(\boldsymbol{\theta}) = \frac{\Sigma^n(\boldsymbol{\theta}) - \overline{\Sigma^n}}{\overline{\Sigma^n}},
\end{align}
where $\Sigma^n(\boldsymbol{\theta})$ is the surface density at position $\boldsymbol{\theta}$ and $\overline{\Sigma^n}$ is the mean surface density of the $n$-th shell for the given cosmology, which we evaluate directly from the shell. 

To exploit the full-sky lightcones, the equation for the lensing potential is solved in spherical harmonics space \citep[e.g.][]{hu2000weak,price2021sparse}. The convergence at plane $n$ is related to the lensing potential using the spherical harmonics coefficients of the convergence ($K^n_{\ell m}$) and lensing potential $(\psi^n_{\ell m})$ as
\begin{align}\label{eqn:lensing_pot_shere_harmon}
    \psi_{\ell m}^n = \frac{-2K_{\ell m}^n}{\ell(\ell+1)},
\end{align}
which can then be used to determine the derivatives necessary to compute the deflection field and shear matrix at each plane. As the mass maps are provided as discrete maps on a HEALPix grid, we need to adopt a strategy for determining the shear matrix and deflection field. For the lowest-redshift shell, the rays can be conveniently `aimed' directly at the pixel centers. Since the photons are deflected here, their paths will, in general, not pass through the center of a pixel in a subsequent shell. Therefore, the magnification matrix and the deflection angle for each ray are evaluated using bilinear interpolation as the weighted average of the four nearest pixels, similar to \citet{shirasaki2015probing}. We have found that the bilinear interpolation introduces some smoothing of the signal, but overall the additional smoothing of the final maps (see below) dominates. Throughout the recurrence relations (Equations~\ref{eqn:mag_mat_recurr}~\&~\ref{eqn:ang_pos_recur}), we determine the lensing potential using Equation~\ref{eqn:lensing_pot_shere_harmon}. The lensing potential is then converted to the deflection field (\textbf{$\alpha$}) and shear matrix ($U_{ij}$) through its first and second-order covariant derivatives, which are used to calculate the magnification matrix ($A$) and the angular position (\textbf{$\beta$}) at each plane, where we take into account the change in basis as the photon gets displaced \citep[][]{becker2013calclens}{}{}.

\subsubsection{Source redshift distribution}\label{sec:nz_source}
\begin{figure}
	\includegraphics[width=\columnwidth]{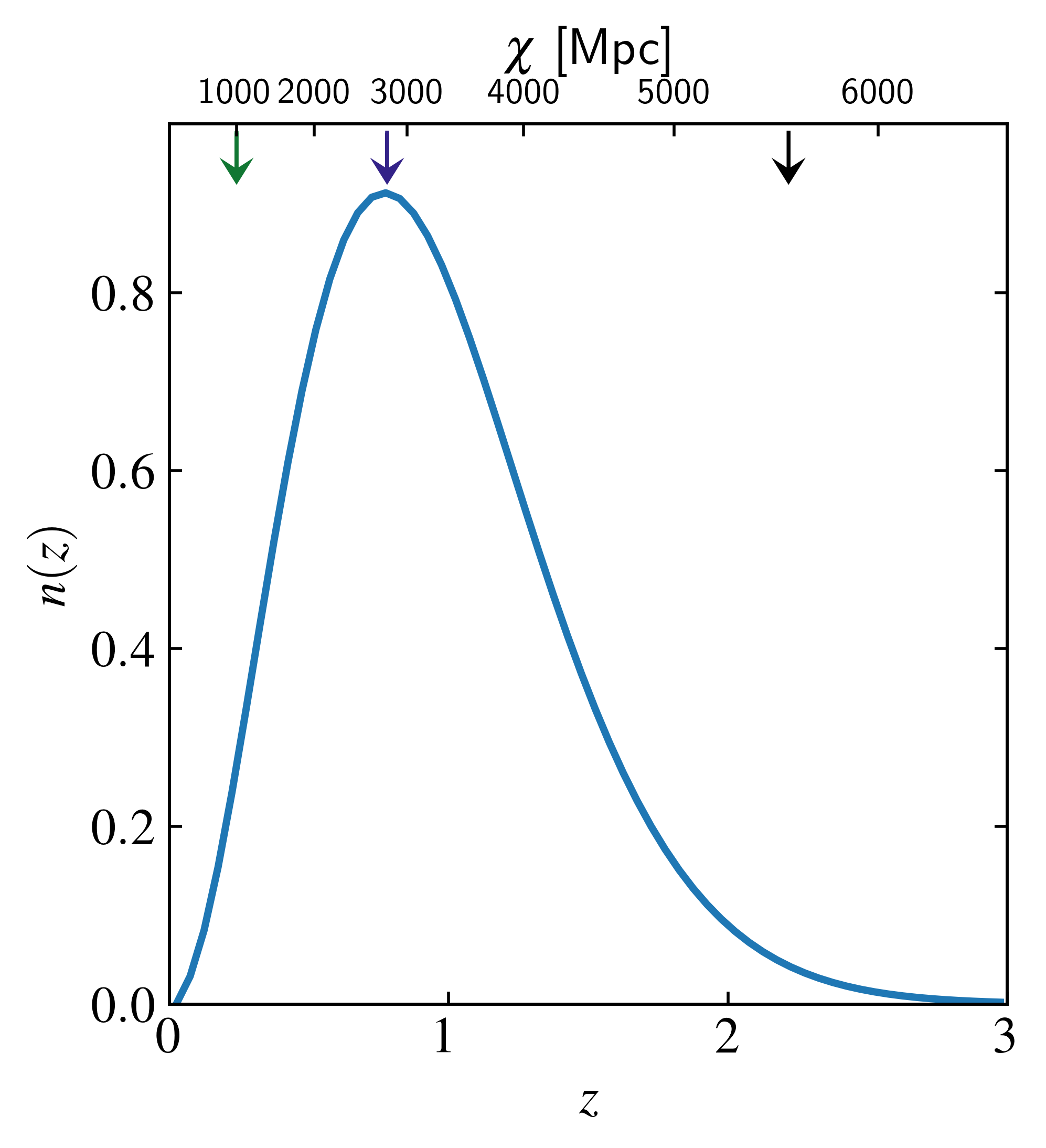}
    \caption{\textit{Euclid}-like source redshift distribution used in the construction of the WL convergence maps. The top axis shows the comoving distance for the fiducial D3A cosmology. The green, blue, and black arrows indicate the box size of the L1 (1\,cGpc), L2p8 (2.8\,cGpc), and  L5p6 (5.6\,cGpc) variations, respectively.}
    \label{fig:source_dist}
\end{figure}

Following the above procedure, the magnification matrix is determined for all rays at all planes. The magnification matrix can then be related to the WL convergence ($\kappa$) via Equation~\ref{eqn:magmat_matrix}. To mimic a signal corresponding to a Stage IV survey, we incorporate a source redshift distribution ($n(z)$) as 
\begin{align}
    \kappa(\theta) = \int_0^{\chi_\mathrm{hor}} \mathrm{d}\chi \, n(z(\chi))\, \kappa(\theta,\chi),
\end{align} 
where the integral runs over the line of sight to the edge of the survey, which in our case reduces to a discrete sum over the shells until $\chi_{\mathrm{hor}} = \chi(z=3)$. Here, we use a simple \textit{Euclid} mock forecast given by \citep[][]{blanchard2020euclid}{}{}:
\begin{align}
    n(z) \propto \bigg(\frac{z}{z_0}\bigg)^2 \exp \bigg[-\bigg(\frac{z}{z_0}\bigg)^{3/2}\bigg],
\end{align}
with $z_0 = 0.9/\sqrt{2}$. The function is normalized such that $\int n(z) \, \mathrm{d}z =1$, and relates to the comoving distance as $n(z)\, \mathrm{d}z = n(\chi)\, \mathrm{d}\chi$. More realistic alternatives to this distribution exist, but for our purposes, this simple distribution suffices as we do not expect minor consistent changes in the redshift distribution to impact the overall baryonic effects. The source redshift distribution is shown in Fig.~\ref{fig:source_dist}. The top axis indicates the comoving distance corresponding to our fiducial cosmology. Additionally, the green, blue, and black arrows indicate the box sizes of the L1, L2p8, and L5p6 variations, respectively. 

\begin{figure*}
	\includegraphics[width=\textwidth]{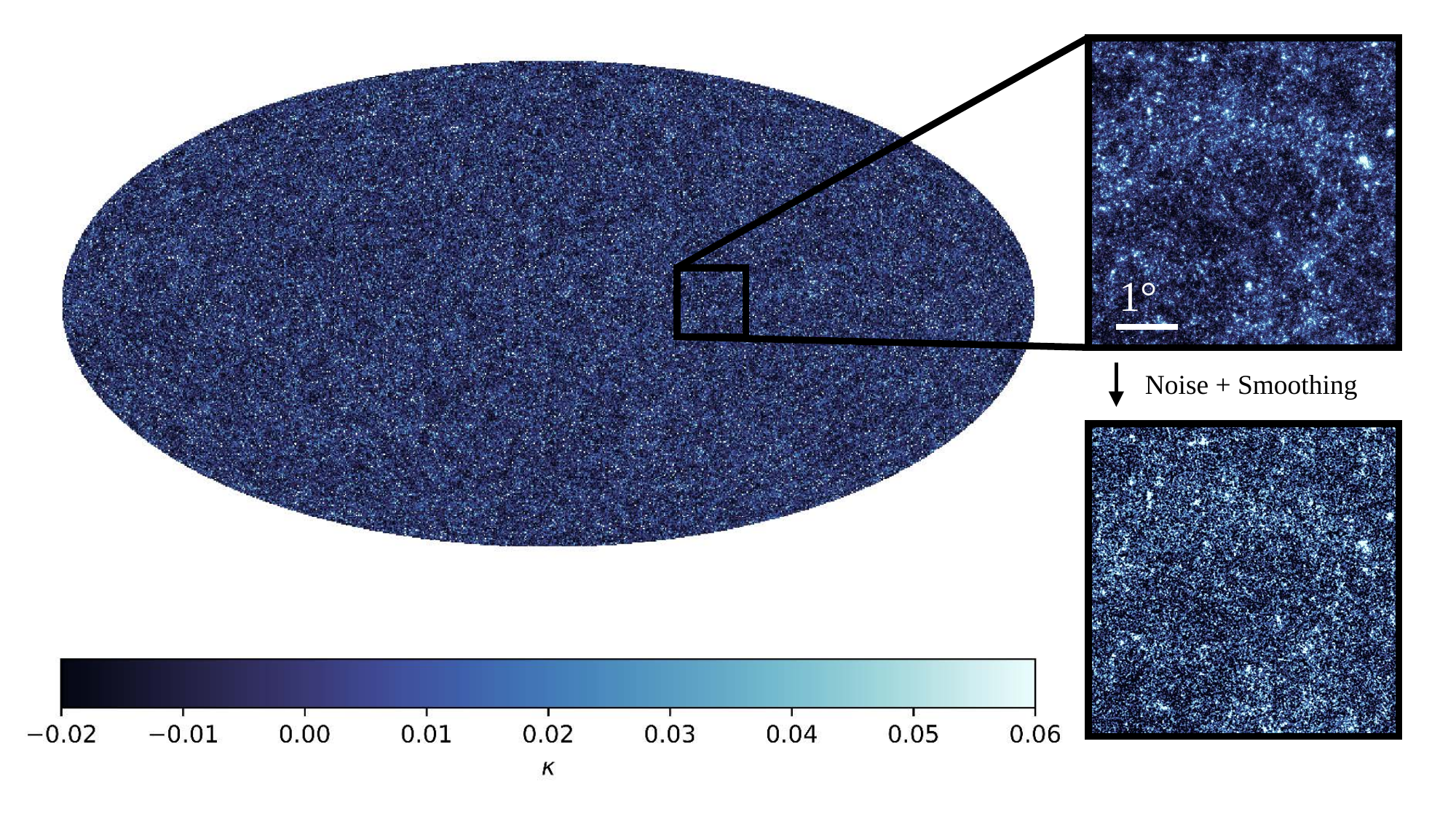}
    \caption{Ray-traced full-sky weak lensing convergence ($\kappa$) map (left) and zoom-ins of a $5\times5$ deg$^2$ region for a \textit{Euclid}-like source redshift distribution as measured by a virtual observer in the hydrodynamical L2p8$\_$m9 run. The full-sky map and top zoom-in correspond to a WL convergence map without smoothing or noise. The bottom zoom-in contains galaxy shape noise and is smoothed with a Gaussian kernel with FWHM = 1 arcmin, both of which are applied before determining the WL peaks. The map is generated using a backward ray-tracing method where the photons that are observed by the virtual observer are deflected by the matter distribution at 60 linearly spaced discrete lensing planes between $z$ = 0 and 3.}
    \label{fig:lightcone}
\end{figure*}

\subsubsection{Smoothing and noise}\label{sec:smoothing_noise}
To mimic an observed signal, we add galaxy shape noise to each pixel of the final map by drawing from a normal distribution with mean $\mu$ and standard deviation $\sigma$ \citep[][]{Kaiser1993noise,lu2021impact}{}{}:
\begin{align}\label{eqn:noise}
   \mathcal{N} \bigg \{ \mu=0,\, \sigma =  \frac{\sigma_\epsilon}{\sqrt{2 n_{\mathrm{gal}} A_{\mathrm{pix}}}} \bigg\},
\end{align}
where $\sigma_\epsilon$ is the rms intrinsic ellipticity of source galaxies, $n_{\mathrm{gal}}$ is the source number density on the sky and $A_{\mathrm{pix}}$ is the pixel area. In our case, we choose $\sigma_\epsilon = 0.26$ and $n_{\mathrm{gal}} = 30$ arcmin$^{-2}$, to mimic the expected signal that \textit{Euclid} will measure \citep[][]{laureijs2011euclid}{}{}. This $\sigma_\epsilon$ value has been measured from \textit{Hubble Space Telescope} (\textit{HST}) images with similar photometric properties as the expected \textit{Euclid} images and is therefore commonly used to model the observed galaxy shape noise \citep[][]{Schrabback2018HST,Martines2019Euclidprep,Euclid2023HOWLS}{}{}. The final maps are smoothed by a Gaussian kernel with a FWHM of 1 arcmin. This was identified by \citet{liu2015cosmology} as an optimal smoothing scale to counterbalance the loss of cosmological information and the minimization of noise. In our case, this corresponds to $\approx2.5$ pixels. 

Fig.~\ref{fig:lightcone} shows a full-sky WL convergence map corresponding to an observer in the L2p8$\_$m9 simulation run. Here, the full-sky map and top $5\times5$ deg$^2$ zoom-in have not been smoothed and no noise has been applied. The bottom zoom-in shows the same area but corresponds to the final WL convergence map where the noise and smoothing have been applied. The highest-valued peaks are visible in both panels but the noise creates spurious low signal-to-noise peaks. 

\subsubsection{Second-order effects}\label{sec:second_order}
Our analysis is somewhat idealized as we ignore several second-order effects such as lens-lens coupling \citep[][]{Bernardeau1997}{}{}, and the fact that the true cosmic shear observable is the reduced shear $g = \gamma / (1-\kappa)$. As explained in \citet{kilbinger2015cosmology}, most of these second-order effects are at least two orders of magnitude smaller than the first-order convergence angular power spectrum. The dominant contribution originates from the reduced shear correction, which contributes up to 10\% of the signal on arc minute scales. However, in our analysis, we assume that the reconstruction of the signal takes into account the reduced shear correction, which can, for example, be done using a quickly converging iteration \citep[][]{Bradac2005}{}{}.

\subsection{Angular power spectra}\label{sec:validation}
\begin{figure}
	\includegraphics[width=\columnwidth]{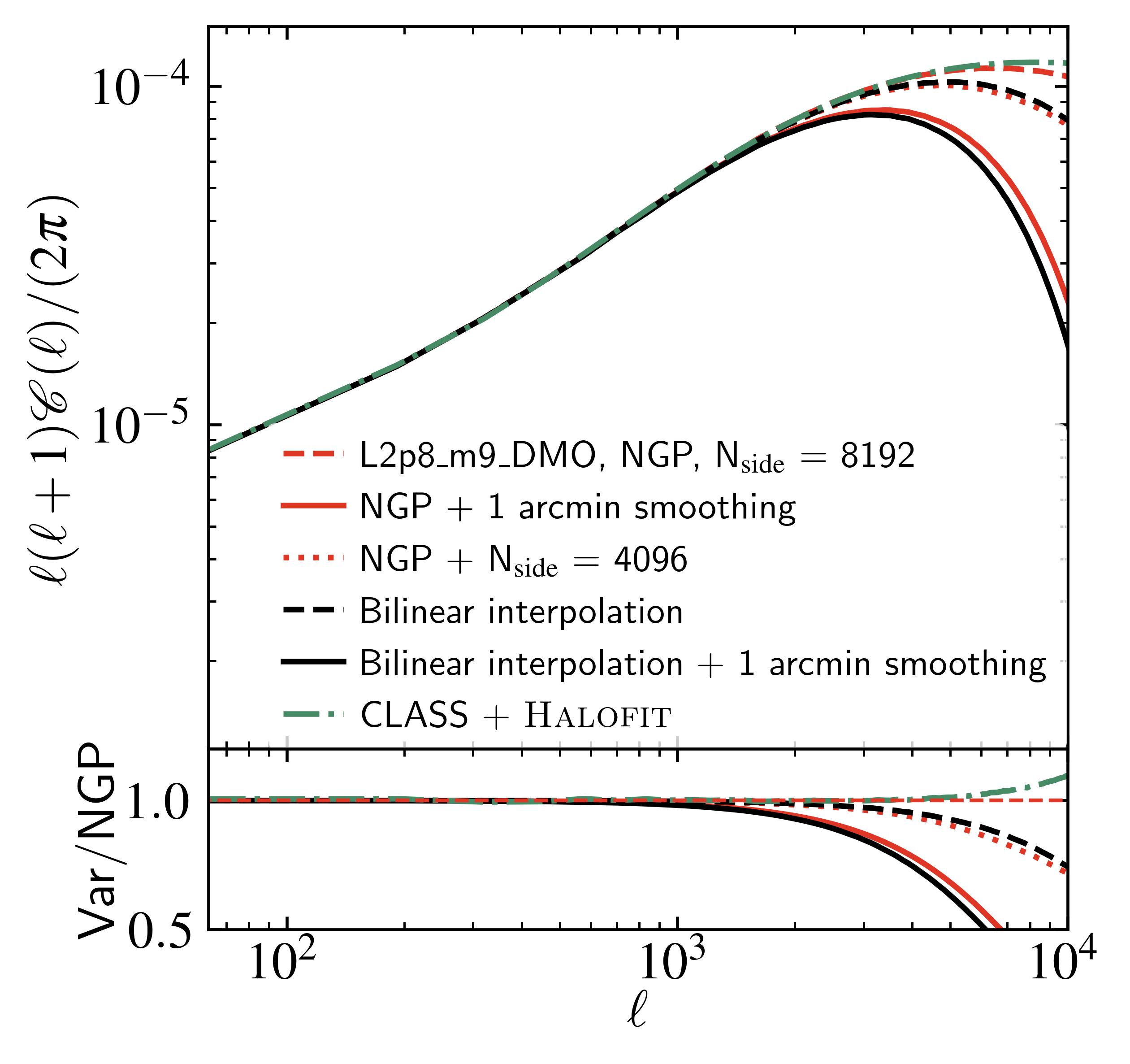}
    \caption{Top: Weak lensing convergence angular power spectrum for variations of smoothing, angular resolution, and interpolation strategy for a single observer in the L2p8$\_$m9$\_$DMO run and the theoretical non-linear prediction from CLASS + \textsc{Halofit}. Bottom: ratios of the different curves in the top panel and the one for nearest gridpoint (NGP) interpolation at $N_{\mathrm{side}}=8192$ (dashed red curve). The 1 arcmin smoothing dominates over any suppression induced by pixelation or ray tracing but differences of a few per cent are still present between different procedures. We use bilinear interpolation at $N_\mathrm{side}=8192$ with 1 arcmin smoothing (solid black curve) as our fiducial map construction method.}
    \label{fig:angpowspec}
\end{figure}

To validate the construction of our WL convergence maps and assess their robustness, we compute the angular power spectrum, $\mathcal{C}(\ell)$, for the L2p8$\_$m9$\_$DMO run for different choices of HEALPix grid resolution, smoothing, and ray-trace methodology. The angular power spectra are computed from the WL convergence map using the HEALPix \textsc{anafast} routine and are shown in Fig.~\ref{fig:angpowspec}. We compare to the prediction from \textsc{Halofit} \citep[][]{Takahashi2020Halofit}{}{} using the CLASS code \citep[][]{blas2011cosmic}{}{}. \textsc{Halofit} is a commonly-used fitting formula to the (non-linear) three-dimensional matter power spectrum ($P_\mathrm{m}$) based on high-resolution $N$-body simulations. We relate the matter power spectrum to the angular power spectrum using \citep[][]{Limber1953approx,Loverde2008limber}{}{}:
\begin{align}
    \mathcal{C}(\ell) = \int_0^{\chi_{\mathrm{hor}}} \mathrm{d} \chi\, \frac{W^2(\chi)}{\chi^2} \,P_{\mathrm{m}}\bigg(\frac{\ell+1/2}{\chi},\, z(\chi)\bigg),
\end{align}
where we assume the fiducial D3A cosmology. $W(\chi)$ is the weak lensing kernel which is given by \citep[][]{Kaiser1992}{}{}:
\begin{align}
    W(\chi) = \frac{3 H_0^2 \Omega_\mathrm{m}}{2c^2}\, \chi \,  (1+z(\chi)) \int_\chi^{\chi_{\mathrm{hor}}} \mathrm{d}\chi' \, n(\chi')\, \frac{\chi'-\chi}{\chi'}.
\end{align}

The \textsc{Halofit} prediction is given by the dash-dotted green curve in Fig.~\ref{fig:angpowspec}. The dashed red curve corresponds to a ray-traced convergence map where the quantities that are determined for each photon at every shell (i.e. the deflection angle and shear matrix), were determined using the value of the nearest gridpoint (NGP) for each ray. In this case, there is excellent agreement between the ray-traced angular power spectrum and the theoretical prediction up to $\ell \approx 6000$. The deviation at smaller scales (larger $\ell$) is a result of the pixelation of the HEALPix grid. Also, as illustrated by \citet{Upadhye2023}, the \textsc{Halofit} prediction deviates from FLAMINGO by a few per cent at small scales. 

The dotted red curve corresponds to the same map as the dashed red curve but downsampled to a lower resolution of $N_{\mathrm{side}}$ = 4096, corresponding to an angular resolution of 0.86 arcmin. As expected, the deviation from the high resolution and \textsc{Halofit} predictions begins at smaller $\ell$ and it is greater in magnitude when the HEALPix grid has a coarser resolution. The dashed black curve corresponds to the case where the quantities at each shell are determined using bilinear interpolation. Because the quantities are determined as a weighted average of the 4 nearest neighbors on the two-dimensional grid, some de facto smoothing at the pixel scale is applied at each shell. We find that this introduces some loss of power on small scales, beyond $\ell \approx 2\times10^3$, to a similar degree as downsampling the resolution of the HEALPix map to $N_{\mathrm{side}} = 4096$. Both the loss of power on small scales due to the smoothing introduced by the interpolation, and that due to the pixelation of the HEALPix grid, are several times smaller than the suppression introduced by smoothing the final maps with a Gaussian kernel with an FWHM of 1 arcmin, as illustrated by the solid red and black curves for NGP and bilinear interpolation, respectively. 

Over all scales, smoothing dominates all the choices that were made in the construction of the maps. However, there is still some additional power loss when using bilinear interpolation compared to NGP interpolation. At $\ell = 2000$ ($\theta \approx 5$~arcmin), the difference between the interpolation strategies is 1\% and it increases for smaller scales. We nevertheless choose to apply bilinear interpolation, which is also used in the ray-tracing method of \citet{hilbert2009ray} and which was shown by \citet{hilbert2020accuracy} to have a peak distribution for peaks with SNR\,<\,6 that agrees to within 5\% with codes that either use higher resolution equidistant cylindrical projection pixelization or determine the convergence signal on the fly \citep[][]{Muciaccia1997,Barreira2017,Fabbian2018}{}{}. \citet{FerlitoMTNGconvergence} have studied the convergence with the pixel size at the same smoothing scale that we use and they find that the distributions are suppressed by up to a couple per cent compared to maps with a higher resolution. We do not expect these few per cent differences to impact our main conclusions as we consistently apply the same approach to all simulation variations. Thus, we still get robust estimates of the differences in number densities of peaks. However, to eventually compare to observations, it might be necessary to use HEALPix maps with a higher angular resolution, whilst sticking to a 1 arcmin smoothing, to avoid sensitivity to the interpolation scheme and to avoid sensitivity to the discretization of the HEALPix grid. 

\section{Results}\label{ch:results}
In this section, we start by quantifying the level of numerical convergence in terms of resolution and box size, and the effect of cosmic variance. We then compare the signals from the different baryonic physics models and the runs with different cosmological models. Finally, we study the separability of the cosmological and baryonic impact. We quantitatively and qualitatively compare our results with previous studies. In our analysis, we quantify the distribution of peaks in the WL convergence maps as a function of their $\kappa$ value. A peak is defined as any pixel that has a value larger than those of the 8 closest pixels on the HEALPix grid. We use the number density of peaks to study the peak distributions. We define the number density of peaks ($\mathrm{d}n / \mathrm{d}\kappa$) as the number of peaks per square degree divided by the convergence bin size \footnote{Although the quantity is independent of bin size, for reference, we state the bin edges rounded to three decimal places [-0.100, -0.085, -0.069, -0.054, -0.040, -0.025, -0.020, -0.015, -0.010, -0.004, 0.001, 0.006, 0.011, 0.016, 0.021, 0.026, 0.032, 0.037, 0.042, 0.047, 0.052, 0.057, 0.062, 0.068, 0.073, 0.078, 0.083, 0.101, 0.153, 0.204, 0.256, 0.308, 0.323, 0.500, 0.563, 1.000]}.

\subsection{Numerical convergence}\label{sec:num_convergence}
\begin{figure}
	\includegraphics[width=\columnwidth]{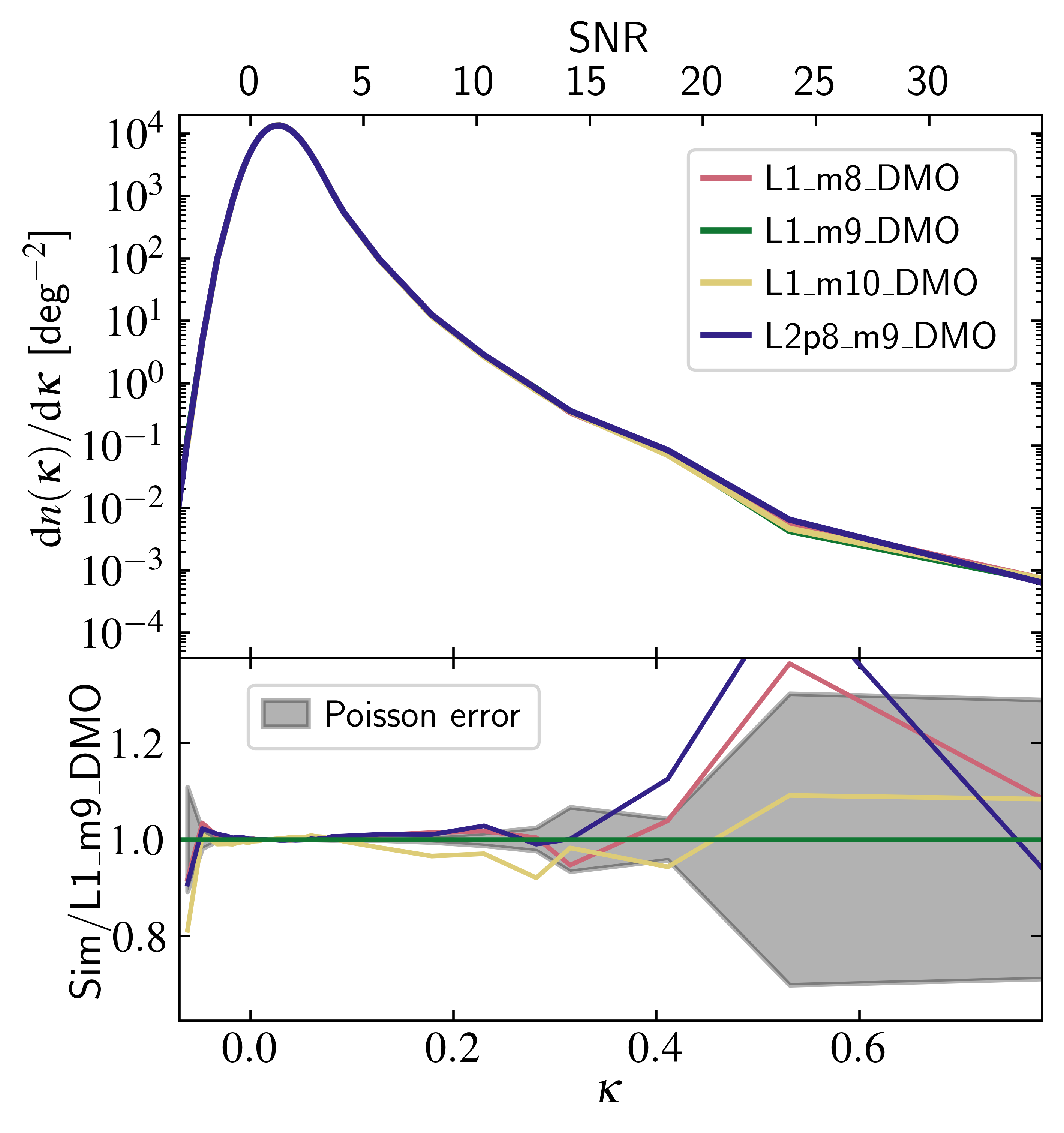}
    \caption{Top: number density of WL peaks for the observers in the L1$\_$m8, L1$\_$m9, L1$\_$m10, and L2p8$\_$m9 DMO runs, which differ in terms of resolution and/or box size. The L2p8$\_$m9$\_$DMO signal is the mean of the 8 observers in the box. The SNR = $\kappa/\sigma$ (top x-axis) is computed from the standard deviation of a smoothed noise realization. Bottom: ratio to the L1$\_$m9$\_$DMO run. The shaded region indicates the Poisson error for L1$\_$m9$\_$DMO. The numerical convergence of the fiducial (m9) resolution and box size (L1), which was used for all the baryonic and cosmology variations, is excellent up to $\kappa =0.3$ and adequate for larger values, where the Poisson error gets larger.}
    \label{fig:mass_res}
\end{figure}

As discussed in Section~\ref{sec:baryonic_variations}, all cosmological and baryonic feedback variations are carried out in L1 boxes at the m9 resolution. Therefore, we first quantify the numerical convergence of the fiducial L1$\_$m9 run with respect to box size (L) and resolution (m). To isolate convergence effects from baryonic effects, we study the numerical convergence in the accompanying DMO$+\nu$ runs, where the convergence of results is well established and understood. To this end, the top panel in Fig.~\ref{fig:mass_res} shows the number density of WL peaks as measured for the observers in L1$\_$m8$\_$DMO (red), L1$\_$m9$\_$DMO (green), L1$\_$m10$\_$DMO (yellow), and the mean of the 8 observers in L2p8$\_$m9$\_$DMO (blue). The high-resolution run has a $64\times$ higher mass resolution than the low-resolution run. The top axis of the figure shows the SNR = $\kappa/\sigma$, where $\sigma$ is the standard deviation of the smoothed galaxy shape noise map (Equation~\ref{eqn:noise}). The bottom panel shows the ratio of the distribution of each model relative to that of L1$\_$m9$\_$DMO. The Poisson error for L1$\_$m9$\_$DMO is indicated by the shaded area in the bottom panel. (The noise does not increase monotonically for larger $\kappa$ as the bin size is not kept constant in order to decrease the noise in large-$\kappa$ bins.) 

The differences between the different resolution L1 runs, which have the same initial phases and are thus not impacted by cosmic variance, can be used to assess the numerical convergence with resolution. The agreement between the fiducial run and the higher resolution run (L1$\_$m8$\_$DMO), as well as with the larger box size run (L2p8$\_$m9$\_$DMO), does not exceed 0.5 (2) \% up to $\kappa = 0.1$ (0.2) and is excellent up to $\kappa \approx 0.4$, illustrating that the measurement is well converged in this regime. These findings are consistent with \citet{FerlitoMTNGconvergence} who studied the numerical and angular resolution convergence of the number density of WL peaks. For larger $\kappa$ values the deviations between the runs become larger than 10\%. However, the number of peaks in this regime also drops and the distribution thus becomes dominated by Poisson noise, as shown by the gray-shaded region. Therefore, we do not expect the peaks in this regime to be cosmologically informative in practice. The lower mass resolution (L1$\_$m10$\_$DMO) shows a lower number of peaks for $\kappa$ $> 0.1$, indicating the number density of WL peaks is not yet converged for resolutions lower than the fiducial m9. As the initial phases of the L1 runs are the same, the same haloes exist in these simulations. At $\kappa$ $\gtrsim 0.4$, the variations between the number densities of the intermediate- and high-resolution runs are likely due to changes in the masses of individual haloes as the positions of the haloes should be unaltered, leaving the WL kernel unchanged. The L2p8$\_$m9$\_$DMO run has different initial conditions and is thus a different realization of the same Universe. The comparison with this run therefore suffers from cosmic variance, which we quantify in the next section. The L1$\_$m8$\_$DMO and L2p8$\_$m9$\_$DMO simulations show a slightly better degree of convergence than the L1$\_$m9$\_$DMO run. Especially up to $\kappa = 0.2$, the larger box and higher mass resolution runs agree almost perfectly. However, this agreement is fortuitous, because it is the result of box size and resolution compensating each other. The comparisons between the resolutions or box sizes illustrate that the signal for L1 and m9, which are used for all cosmological and baryonic variations, is well converged in the regime $\kappa \lesssim 0.4$. Unlike larger $\kappa$ values, this regime is not impacted significantly by Poisson errors and cosmic variance (as shown in the next section), and will thus be most informative when inferring cosmological constraints.

\subsection{Cosmic variance}\label{sec:cosmic_var}
\begin{figure}
	\includegraphics[width=\columnwidth]{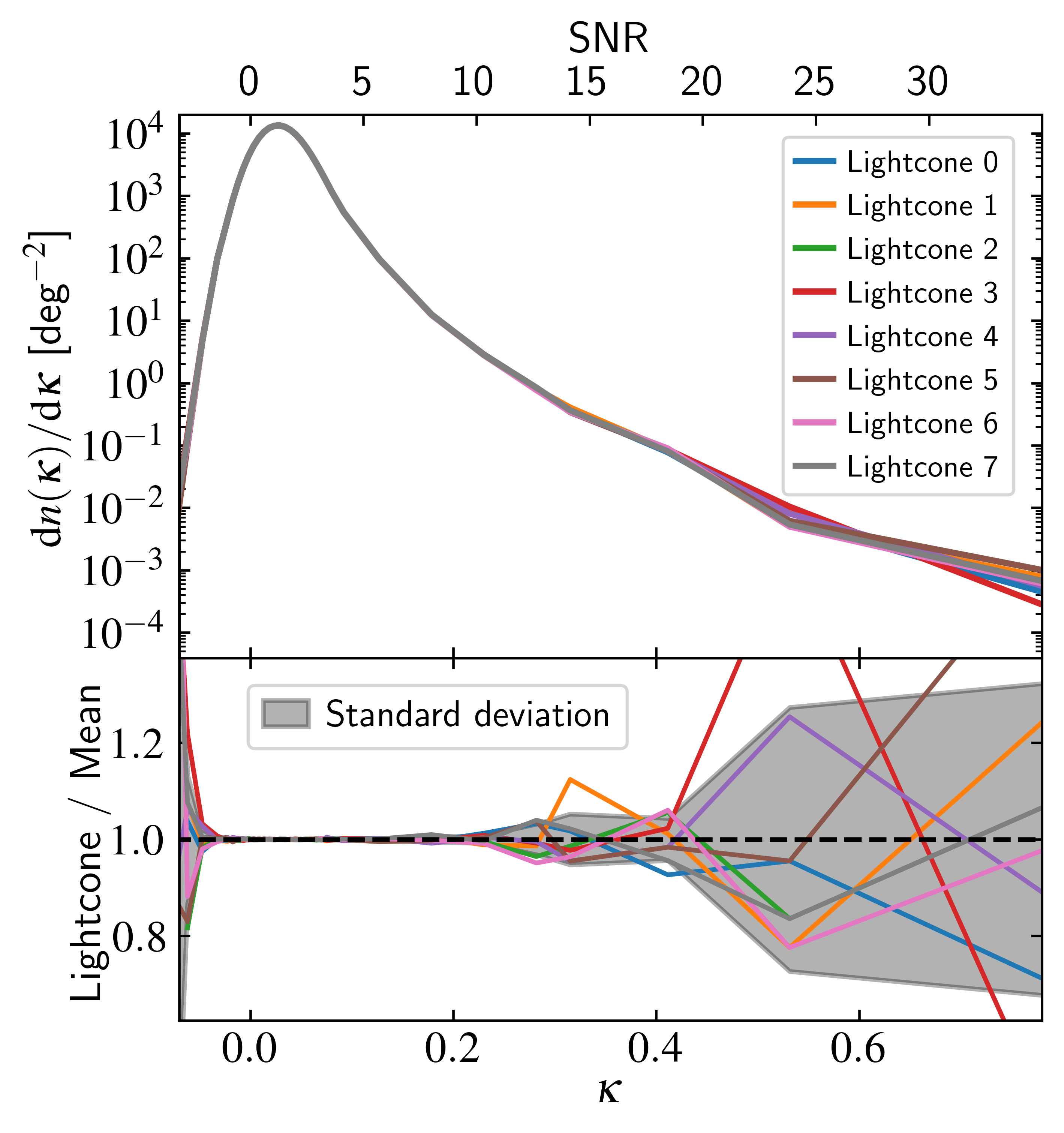}
    \caption{Top: number density of WL peaks for 8 different virtual observers in the L2p8$\_$m9$\_$DMO simulation. The observers were placed at the coordinates ($\pm L/4$,$\pm L/4$,$\pm L/4$), where $L = 2.8$~cGpc is the simulation box size. Bottom: Ratio with the mean of the 8 observers. The gray shading shows the standard deviation of the 8 observers. Over the entire range of $\kappa$, the cosmic variance is comparable to the Poisson errors. At $\kappa > 0.4$, the uncertainty due to cosmic variance increases rapidly, which will be the limiting factor for any inference based on the number density of WL peaks.}
    \label{fig:cosmic_var}
\end{figure}

Using the 8 independent observers within the L2p8$\_$m9$\_$DMO run, we can directly test the impact of cosmic variance on the measured statistics. The top panel of Fig.~\ref{fig:cosmic_var} shows the number density of peaks for each of the 8 lightcones. The bottom panel shows their ratios with their mean. The standard deviation of the ratio is indicated by the gray shaded region. For positive WL convergence values up to $\kappa = 0.1 (0.2)$, the distributions agree to within 0.2 (1)\% precision. Up to $\kappa \approx 0.4$, the difference is not greater than 10\%, after which the degree of variation increases sharply. For $\kappa > 0.4$, the number density is so low that there exist fewer than 10$^{2}$ peaks within a convergence bin on the entire sphere. As the WL convergence signal is most sensitive to overdensities roughly halfway in between the observer and the source galaxy, the exact configuration of the observers with respect to the most massive haloes in the simulation will determine the number of peaks in the highest-$\kappa$ bins \citep[][]{kilbinger2015cosmology}{}{}. This is reflected in the distributions as the variance increases for larger WL convergence values. 

Based on the comparisons in this and the previous section, we find that our measurements are robust up to at least $\kappa = 0.4$. For larger WL convergence values, the difference increases but both the Poisson noise as well as the uncertainty due to cosmic variance start to dominate the signal, which will be the limiting factor for any WL survey using the number density of high-$\kappa$ WL peaks. The comparison in this and the previous section shows that the most informative regime for the number density of WL peaks is $\kappa \approx 0.1-0.4$. The $\kappa >0.4$ regime suffers from cosmic variance and Poisson noise and $\kappa < 0.1$ is impacted by the smoothing and noise, as is discussed more in the next section. Comparing the cosmic variance effect to the Poisson uncertainty in Fig.~\ref{fig:mass_res}, we see that over the entire $\kappa$ range, the two are of similar magnitude. 

\subsection{Comparison of fiducial dark matter only and hydrodynamical models}
\begin{figure}
	\includegraphics[width=\columnwidth]{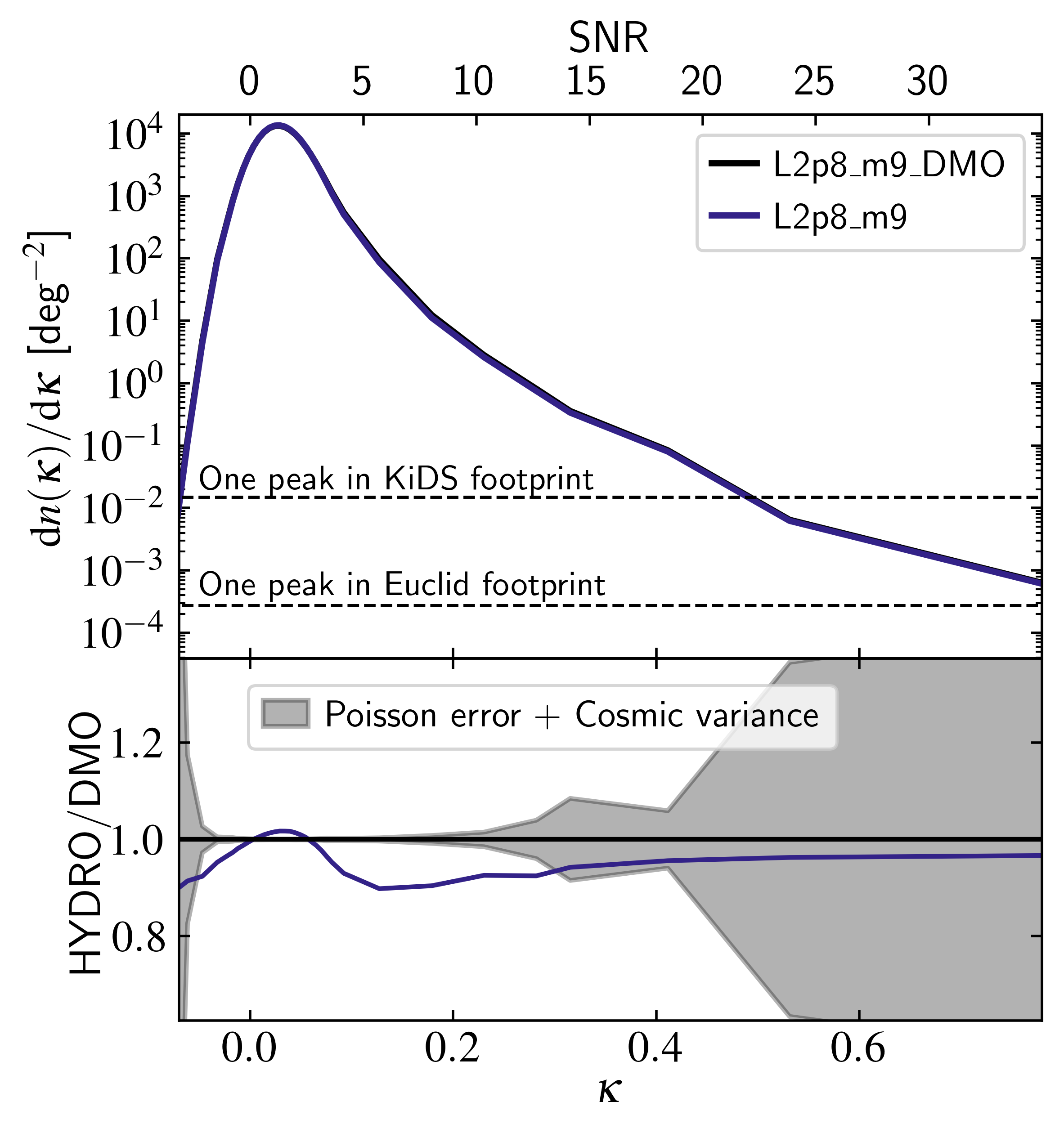}
    \caption{Top: mean number density of WL peaks measured by the 8 virtual observers in the DMO and cosmological hydrodynamical L2p8$\_$m9 runs with the same cosmology and initial phases. The observers are located at the same positions in the simulations and we apply the same random rotations to their lightcone shells. The intersection of the dashed horizontal lines with the number densities indicates the $\kappa$ regime where we expect to measure only a single peak with a larger $\kappa$ value in the KiDS or \textit{Euclid} footprints. Bottom: ratio of the mean number density of peaks for the observers in the hydrodynamical run to the mean in the DMO run. The gray shaded area indicates the quadrature sum of the estimate Poisson error and cosmic variance. The fiducial baryonic impact is largest at $\kappa \approx 0.1$, showing a 10\% suppression of the number density of WL peaks. The suppression due to baryons decreases almost monotonically for larger $\kappa$.}
    \label{fig:dmo_vs_hydro}
\end{figure}

In this section, we explore the differences between the number density of WL peak distributions measured in a DMO and hydrodynamical run. To this end, we compare the mean of the measured signal of the 8 virtual observers in the L2p8$\_$m9 and L2p8$\_$m9$\_$DMO runs. Their mean number density of peak distributions are shown in the top panel of Fig.~\ref{fig:dmo_vs_hydro} in blue and black, respectively. As the initial phases, observer positions, and applied rotations are the same, any difference is a direct measurement of the impact of baryonic physics on the WL convergence peak count distribution. The bottom panel shows the ratio of the mean of the hydrodynamical run compared to that of the DMO distributions. The dashed horizontal lines in the top panel indicate the regime at which we expect to measure only one peak in the \textit{Euclid} footprint \citep[$\approx 15000$~deg$^2$;][]{laureijs2011euclid}{}{} and the KiDS footprint \citep[$\approx 1550$ deg$^2$;][]{Kuijken2019KiDS}{}{} with a larger WL convergence value. As \textit{Euclid} will cover $\sim1/3$rd of the sky, close to the entire convergence regime can be probed. At small convergence values, $0 \lesssim \kappa \lesssim 0.05$, the baryonic physics enhances the number density of WL peaks by a few per cent. In Appendix~\ref{app:smooth_noise} we explore the impact of the applied smoothing and noise and illustrate that without the smoothing and noise, the hydrodynamic run shows a stronger enhancement of WL peaks around $\kappa = 0$. The comparison shows that the entire $\kappa$ regime is impacted by the smoothing and noise and illustrates that for an actual inference, a more dedicated study that models these effects properly must be carried out. 

At $\kappa \gtrsim 0.1$, there is a clear difference between the models, where we see a larger number of peaks in the DMO run. At $\kappa \gtrsim 0.4$, the number of peaks becomes very small and although the trend is still clear, this regime is dominated by Poisson noise and cosmic variance, as shown in Sections~\ref{sec:num_convergence} \& \ref{sec:cosmic_var}. The gray-shaded area corresponds to the quadrature sum of the Poisson error and cosmic variance as estimated for L2p8$\_$m9$\_$DMO, which will thus dominate over the baryonic impact for $\kappa > 0.4$.

We can understand the baryonic suppression of the peak distribution by considering differences for individual haloes between the hydrodynamical and DMO runs. Baryonic feedback will cause gas to be expelled from a halo causing the halo to be less massive \citep[e.g.][]{Velliscig2014,Bocquet2016}{}{}, which directly decreases its lensing potential. Also, when gas is expelled early in a halo's lifetime, it will be less massive and consequently have a less deep gravitational potential and thus attract less matter over its entire evolution, increasing the mass difference compared to the DMO run \citep[][]{Stanek2009,Cui2012}{}{}. \citet{Debackere2022} compared the halo masses of matched haloes in BAHAMAS and its DMO counterpart and showed that for haloes of mass $M_{\mathrm{200m}} \approx [10^{13}-10^{14}]\,\mathrm{M_\odot}$, the halo mass in the hydrodynamical simulation is $\sim 10 \%$ lower than in the DMO run. As the WL convergence signal is directly proportional to the overdensity (Equation~\ref{eqn:conv_at_shell}) and thus mass, we expect to observe weaker WL signals in the hydrodynamical run. The reduction in mass in hydrodynamical runs is directly reflected in the observed number densities, where we see a suppression of a similar magnitude. As shown by \citet{Debackere2022}, for even more massive haloes ($M_{\mathrm{200m}} \gtrsim 10^{14}~\mathrm{M_\odot}$), the feedback is not strong enough to effectively remove large amounts of gas from the haloes, and the mass difference between the haloes in different runs decreases. The same patterns are visible in the FLAMINGO HMFs reported in Fig.~20 of \citet{schaye2023flamingo}, where compared to the DMO counterpart, the HMF is suppressed in the intermediate halo mass regime ($M_{\mathrm{200m}} \approx [10^{13}-10^{14}]\,\mathrm{M_\odot}$) by 10-20\%, with larger differences for less massive haloes, but for the most massive haloes ($M_{\mathrm{200m}} \approx 10^{15}\,\mathrm{M_\odot}$) the suppression vanishes. We observe a similar effect as we see a decreasing suppression of the number density of WL peaks for $\kappa >0.1$. Although the differences in halo mass and WL peak abundances agree quantitatively, we cannot conclude that haloes of specific masses are primarily responsible for the peaks. The WL convergence signal is also sensitive to the orientation of the observer with respect to the haloes and a peak with a larger $\kappa$ value will thus not necessarily correspond to a more massive halo. However, in general, peaks with a larger $\kappa$ value are more likely to originate from more massive haloes so qualitatively we do expect to see less suppression for larger $\kappa$ values as we understand these peaks to originate from a single halo along the line of sight and the mass of the most massive haloes should be similar in the DMO and hydrodynamical runs \citep[][]{yang2011cosmological}{}{}. We aim to study the contribution of haloes of a specific mass range in future research.

We now compare our results with those from previous studies. We note that differences between our and previous analyses may arise because of different baryonic feedback implementations, source redshift distributions, and WL convergence map construction algorithms, which rely on choices for shape noise, smoothing, and ray tracing. 

\citet{coulton2020weak} reported on the impact of baryons on peak counts for a Rubin-like inference using the BAHAMAS simulations. Their analysis, which only extends to SNR\,=\,6, also includes noise and smoothing, but their smoothing scale is twice as large as ours. In this convergence regime ($\kappa \lesssim$ 0.1), which we found was most impacted by the applied noise and smoothing, they report a baryonic suppression that is roughly half of the suppression we measure. The reason for the difference is unclear, as the BAHAMAS runs were calibrated to match the same observables, but it illustrates that the magnitude of the baryonic suppression may depend on the details of the subgrid prescriptions in the simulation as well as the choices made in the construction of the convergence maps. \citet{Osato2021baryons} have carried out a similar analysis, including a full ray-tracing treatment, in Illustris TNG \citep[][]{Springel2018Illustris}{}{} and they too found a suppression of the peak counts. Whereas they considered a single-redshift source sample and their smoothing scale is twice as large, they report a similar suppression factor. We cannot compare our results directly to the ones extracted by \citet{FerlitoMTNGconvergence} from the Millenium-TNG simulation, who studied convergence peaks in a suite of different hydrodynamical simulations, as their analysis only focuses on convergence values up to $\kappa = 0.06$ and they do not apply any observationally-inspired shape noise. In our case, this regime is completely dominated by the applied smoothing and shape noise. 

In this section, we have shown that the fiducial baryonic feedback within the FLAMINGO simulation suite suppresses the number density of WL peaks as measured for a Stage IV WL survey by $10\%$, and that the suppression decreases for larger $\kappa$ values. The suppression can be understood by considering the impact of feedback, which expels gas from haloes and thus decreases their mass and thereby the WL potential.

\subsection{Baryonic variations}
\begin{figure}
	\includegraphics[width=\columnwidth]{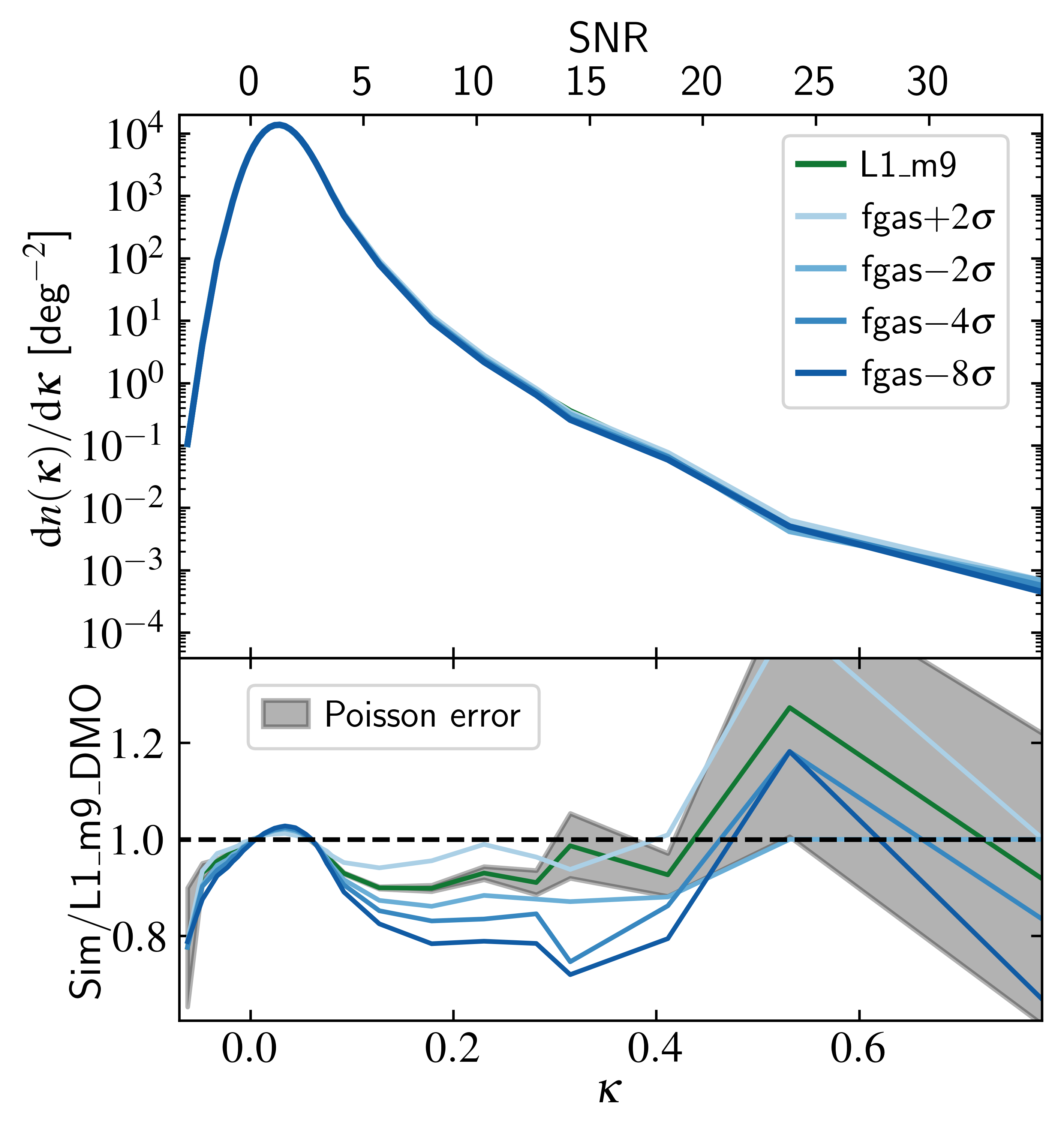}
    \caption{As Fig.~\ref{fig:dmo_vs_hydro} but comparing the runs calibrated to different gas fractions in cluster, as indicated by fgas$-$n$\sigma$ where n is the change in the number of standard deviations compared to the fiducial L1$\_$m9 (green curve) model, to L1$\_$m9$\_$DMO. Models with stronger feedback, indicated by an increasingly darker blue color, have progressively stronger suppressed number densities of WL peaks.}
    \label{fig:gas_fracs}
\end{figure}

Next, we compare the number density of peaks distributions in the variations that were calibrated to different observables. First, we compare the models that were calibrated to different gas fractions in clusters. We stress that these variations were run in similar boxes (L1) with the same resolution (m9), have identical subgrid feedback implementations, were run assuming the same cosmology, have the same initial conditions and the virtual observer was placed at the same position. The only difference between the runs is the value of 4 subgrid parameters, as described in Section~\ref{sec:baryonic_variations}, which were chosen to change the resulting gas fraction ($f_{\mathrm{gas}}$) in clusters by a set amount. The number density of peaks for the runs with different gas fractions are shown in Fig.~\ref{fig:gas_fracs}. Again, at $\kappa < 0.1$, the differences between the distributions are washed out due to the applied smoothing and noised.

In general, compared to the fiducial L1$\_$m9, the lower (higher) gas fraction models correspond to simulations in which more (less) baryons have been evacuated from their haloes, as stronger (weaker) baryonic feedback is present. As the runs have the same initial conditions, the same haloes exist at the same positions in the simulation and comparisons between the runs are not affected by cosmic variance. We therefore include only the estimate of the L1$\_$m9 Poisson noise in the lower panel. Any difference between the variations is a direct result of the halo mass differences as the WL kernel does not change. Fig.~\ref{fig:gas_fracs} shows that the models with lower gas fractions, which are indicated by an increasingly darker blue color, have progressively smaller number densities of WL peaks in the intermediate convergence regime ($\kappa \approx [0.1,0.4]$), with the differences increasing up to $\kappa \approx 0.2$. We can understand the lower gas fractions as being the result of stronger feedback. Stronger feedback leads to more gas being expelled from the centers of haloes, and thus also to smaller overdensities. We therefore expect to see a stronger suppression of the WL peak counts for models calibrated to lower gas fractions. The hierarchy in gas fraction models agrees with the HMFs reported by \citet{schaye2023flamingo}, where they show that in the stronger feedback models, the HMF is increasingly more suppressed in the $10^{13}\,\mathrm{M_\odot} \lesssim M_{\mathrm{200m}}  \lesssim 5\times10^{14}\,\mathrm{M_\odot}$ regime. The number of haloes in this mass regime is smaller in the models with lower gas fractions and we thus expect there to be fewer high-valued WL convergence peaks too.

\begin{figure}
	\includegraphics[width=\columnwidth]{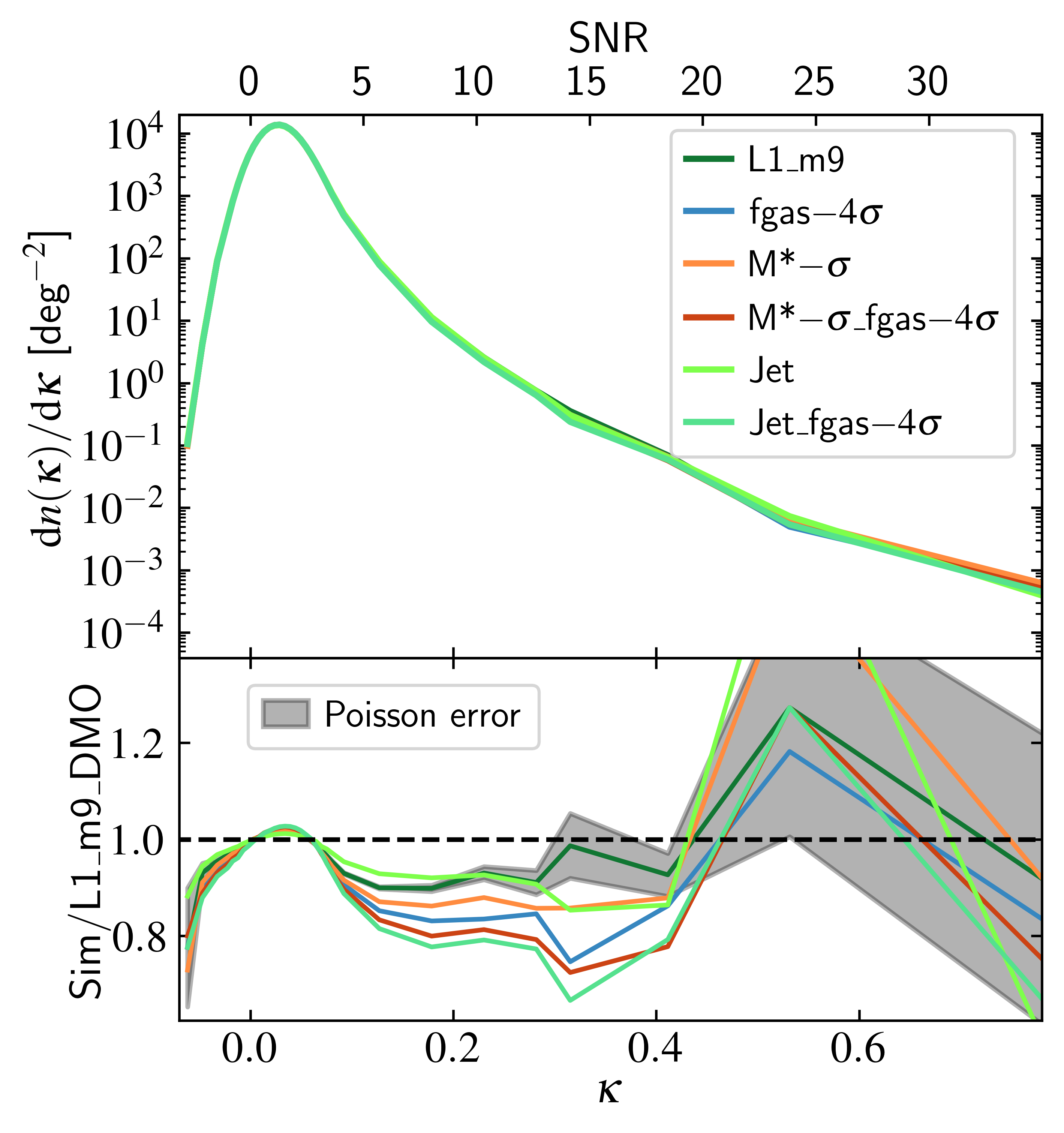}
    \caption{As Fig.~\ref{fig:dmo_vs_hydro} but comparing the baryonic feedback models varying the galaxy stellar mass function (M*), AGN feedback prescription (Jet), gas fraction in clusters (fgas), or a combination of these, to L1$\_$m9$\_$DMO. Calibration to lower galaxy stellar mass functions leads to suppressed number densities of WL peaks. Variations in AGN subgrid prescriptions can lead to differences in WL peak counts not captured by the gas fraction in clusters.}
    \label{fig:feedback_vars}
\end{figure}

Next, we compare the remaining baryonic model implementations in Fig.~\ref{fig:feedback_vars}.
The L1$\_$m9 (dark green) and Jet (light green) models, which differ in the subgrid implementation of AGN feedback (thermal or jet) but have been calibrated to the same observables, show a different trend in their number density profiles. Whereas the baryonic suppression for L1$\_$m9 is largest (10\%) at $\kappa = 0.1$, for model Jet it increases to $\approx 20\%$ at $\kappa = 0.4$. In contrast, the fgas$-4\sigma$ (blue) and Jet$\_$fgas$-4\sigma$ (green) variations, which are each calibrated to the same observables but to a different gas fraction than the fiducial run, show a similar trend but a systematic difference of $\approx 10$\% for $0.1 < \kappa < 0.4$, where the jet feedback model predicts smaller number densities. The difference between the two comparisons illustrates that the baryonic suppression of the number density of WL peaks cannot be expressed only in terms of the gas fraction in clusters, but also depends on how the astrophysical feedback is implemented and the gas distribution is reshaped, as jet feedback can potentially move mass further out \citep[e.g.][]{Federrath2014}{}{}. In this case, we can only partly understand the difference between the runs based on the HMFs. The HMFs of these runs in \citet{schaye2023flamingo}, show other differences. For halo masses of $10^{13}\,\mathrm{M_\odot} \lesssim M_{\mathrm{200m}}  \lesssim 10^{14}\,\mathrm{M_\odot}$ the thermal and jet models calibrated to the fiducial gas fraction show 5-20\% differences, with L1$\_$m9 having a lower HMF, whereas the fgas$-4\sigma$ models do not differ by more than 10\% from each other. At larger halo masses, up to $M_{\mathrm{200m}}\approx 5\times10^{14}\,\mathrm{M_\odot}$, the difference between L1$\_$m9 and Jet vanishes whereas the difference between the fgas$-4\sigma$ and Jet$\_$fgas$-4\sigma$ models remains of similar magnitude. Possibly, the peaks we measure primarily originate from haloes with $M_{\mathrm{200m}} > 10^{14}\,\mathrm{M_\odot}$, as the difference between the fgas$-4\sigma$ models is larger in that regime. \citet{liu2016origin} have used a halo model to study the origin of WL convergence peaks in the Canada-France-Hawaii Telescope Lensing Survey (CHFTLenS) and found that the highest valued peaks are caused by a single massive halo of mass $ M_{\mathrm{vir}} \approx 10^{15}\,\mathrm{M_\odot}$. It is unclear to what extent haloes of a certain mass contribute to which $\kappa$ values of the WL peaks. We aim to investigate this in future research. 

Next, we compare the models that have been calibrated to the same gas fraction but to different galaxy stellar mass functions. Compared to the fiducial model, we see that the model with a lower galaxy stellar mass function (M*$-\sigma$; orange) also gives a suppressed number density of WL peaks. Within the simulation, to have a lower SMF, stronger feedback is required on galaxy scales. At the same time, the model was calibrated to have the same cluster gas fractions. The overall stronger feedback is reflected in the WL peak counts, as we see a suppression, suggesting the masses of the lenses are smaller than in the fiducial model. Compared to the difference between L1$\_$m9 and M*$-\sigma$, the difference between the fgas$-4\sigma$ and M*$-\sigma\_$fgas$-4\sigma$ models is slightly smaller but in general of similar magnitude and sign, suggesting the effects of galaxy- and cluster-scale feedback are (partly) separable. These differences are consistent with the HMF interpretation and results reported in \citet{schaye2023flamingo}. For $\kappa \gtrsim 0.4$, the regime dominated by Poisson noise and cosmic variance, the differences between the runs are less distinctive and informative. 

Next, we consider the difference in the sensitivity to the two observables to which the FLAMINGO simulations were calibrated. Comparing the gas fraction variations in Fig.~\ref{fig:gas_fracs} to the stellar mass variations in Fig.~\ref{fig:feedback_vars}, we see that M*$-\sigma$ shows a similar suppression as the fgas$-2\sigma$ model, both showing a suppression of $\approx~15\%$ for $\kappa = 0.1-0.4$, suggesting the WL observable is slightly more sensitive to the deviations as set by their current uncertainty in the SMF than the gas fraction in clusters. The M*$-\sigma\_$fgas$-4\sigma$ model shows some additional suppression compared to fgas$-4\sigma$ as the former run seems to fall in between fgas$-4\sigma$ and fgas$-8\sigma$, showing the suppression factors that are caused by lowering the two observables are at least partly independent. However, the suppression is smaller than that shown by the Jet$\_$fgas$-4\sigma$ model, illustrating that different feedback prescriptions can already cause stronger differences than the shift in the SMF.

In summary, the suppression measured in the WL peak abundances between the FLAMINGO baryonic feedback variations is 5-25\% for WL convergence values of $\kappa = 0.1 - 0.4$, with models with stronger feedback showing a larger suppression. The peak abundance is sensitive to the amount of gas that is displaced from haloes and can generally be understood qualitatively by considering the impact of feedback on single haloes and the differences in the HMFs. The exact baryonic suppression of the WL peak counts is sensitive to both the subgrid feedback prescription, as varied between the thermal en jet models, as well as the feedback strength leading to different cluster gas fractions and galaxy stellar mass functions.

\subsection{Cosmology variations}

\begin{figure}
	\includegraphics[width=\columnwidth]{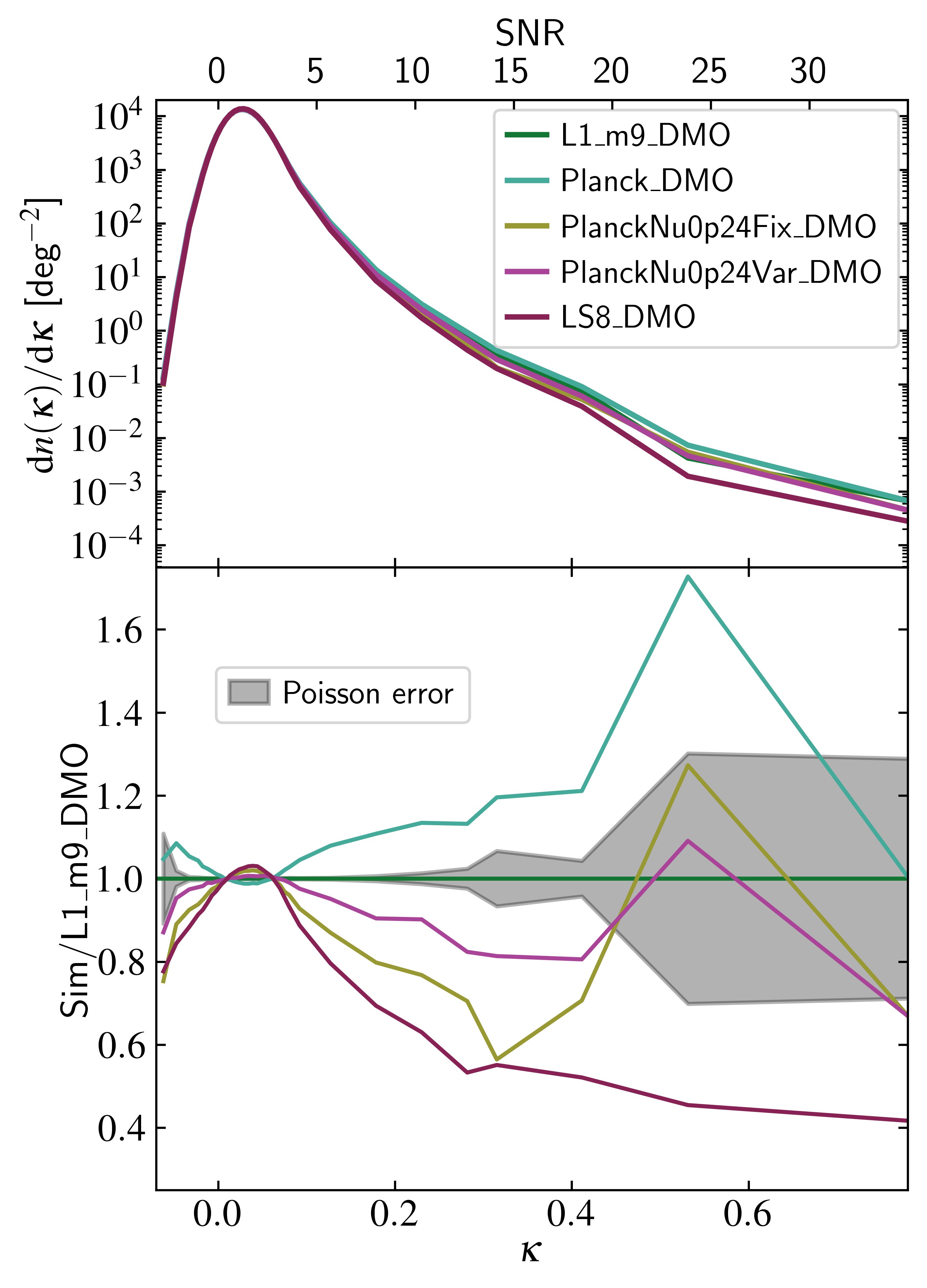}
    \caption{As Fig.~\ref{fig:dmo_vs_hydro} but comparing the cosmology variations in the 1\,cGpc DMO box. The cosmologies are listed in Table~\ref{tab:cosmologies}. The vertical range in the bottom panel is larger than in all similar plots as the cosmology variations, in general, show larger differences than the baryonic feedback variations. The differences can be qualitatively understood from the impact of cosmological parameters (primarily $\Omega_\mathrm{m}$ and $\sigma_8$) on the HMF.} 
    \label{fig:cosmo_vars}
\end{figure}

In this section, we study the effect of varying the cosmology. The top panel in Fig.~\ref{fig:cosmo_vars} shows the number density of the peaks measured for the different cosmology variations in the L1 box. The bottom panel again shows the ratio with the fiducial L1$\_$m9 run. Note that the vertical range on the $y$-axis is larger than that of the previous plots, as the differences are, in general, larger. The most notably different model is the LS8 model. Across the entire $\kappa$ range where the signal is not dominated by smoothing and noise, LS8 has a lower number density of peaks than any of the other models. Here, we can understand the differences between the different distributions by considering the impact of different cosmological parameters on the HMF. The HMF depends strongly on the cosmological parameters $\Omega_\mathrm{m}$ and $\sigma_8$, where a lower $\Omega_\mathrm{m}$ decreases the overall amplitude of the HMF and $\sigma_8$ primarily influences the high mass end of the HMF, moving the exponential cutoff to lower masses for lower values of $\sigma_8$ \citep[e.g.][]{tinker2008toward,Xhakaj2023hmf}{}{}. Compared to the fiducial model, the LS8 model has almost the same value of $\Omega_\mathrm{m}$ but a 6\% lower value of $\sigma_8$. We thus expect fewer massive haloes in the LS8 model. This is reflected in Fig.~\ref{fig:cosmo_vars}, where we see that the number density of WL peaks is the lowest for the LS8 run across the entire intermediate convergence regime ($\kappa \approx [0.1,0.4]$). As shown by \citet{li2019constraining}, for a Stage IV WL survey, WL peaks with SNR\,>\,3 each originate from a single massive halo along the line of sight, and the number of these haloes is thus affected by the lower $\sigma_8$ value. Their SNR value does not directly correspond to our SNR=3 as the noise, smoothing, and redshift distribution are different. Nevertheless, we understand the highest peaks to be caused by a single halo. We therefore expect the number density of (high) WL peaks to be lower in the LS8 run. The deviation from the fiducial model increases for larger convergence values. Following similar reasoning, we can understand why the Planck cosmology model shows the largest number density of peaks. Compared to the fiducial cosmology, the Planck cosmology has larger values of both $\Omega_\mathrm{m}$ and $\sigma_8$. We thus expect the total number of haloes to be larger as well as the cutoff at high halo mass to shift to higher masses, which is reflected in the bottom panel of Fig.~\ref{fig:cosmo_vars}, as the Planck model has larger number densities than the other variations for $\kappa > 0.1$.

Our results agree qualitatively with \citet{coulton2020weak}, who varied the cosmological parameters $\Omega_{\mathrm{m}}$ and $A_\mathrm{s}$. Our analysis, which extends to larger $\kappa$ values, shows that the model with a higher total matter density continues to have an enhanced number density of WL peaks for the entire WL convergence regime. In our case, in the intermediate convergence regime ($\kappa \approx [0.1,0.4]$), the difference increases from 10 to 20\%. In both instances, the model with the lower value of $A_s$, in our case the LS8 model, has a slightly increased number density of WL peaks for SNR $ = 0$ to $3$, whereas for larger values, the runs with the lower values of $A_s$ show a suppression of the number density of WL peaks. Our analysis shows this holds up to the largest values of $\kappa$. 

We now turn to the models with varying neutrino masses. We observe that the number density of peaks corresponding to the heavier neutrino models (Planck0p24Fix and Planck0p24Var), in the intermediate WL convergence regime, is $20-40\%$ lower compared to the Planck model. In cosmology, we understand neutrinos to act as a form of hot dark matter (HDM). Due to their high speeds and weak interactions with regular matter, neutrinos cannot be contained effectively in regions smaller than their free-streaming length. Therefore, they can carry mass away from overdense regions, impeding the growth of clusters \citep[e.g.][]{Lesgourgues2006neutrinos}{}{}. The effect of neutrinos on the HMF has been studied in the BAHAMAS simulation by \citet{mummery2017neutrinos}, who showed that the massive end of the HMF is preferentially suppressed by the free streaming of massive neutrinos. In their comparison, the difference in the HMF in the halo mass regime of [$10^{14}-10^{15}$]\,$\mathrm{M_\odot}$ is $10-20$\% for a similar neutrino mass difference as we consider. Model Planck0p24Fix, for which the values of the cosmological parameters other than the neutrino mass are the same as for model Planck, shows a $10\%$ stronger suppression than the  Planck0p24Var model. Qualitatively, we find similar results to \citet{coulton2020weak}, who studied the peak count distribution with varying cosmological parameters in the $N$-body \textsc{MassiveNus} simulations \citep[][]{Liu2018MassiveNus}{}{}, as we see that more massive neutrinos suppress the WL peak counts. Quantitatively, our suppression is a few factors larger than they report. The most obvious explanation is that we have $\Delta \sum m_\nu c^2 = 0.18$~eV, whereas they have a summed mass difference of 0.1~eV. We thus expect the suppression we measure to be larger than the value they find. Our results also agree quantitatively with \citet{FerlitoMTNGconvergence} who studied peak counts in MTNG variations with varying neutrino mass. They consider two MTNG variations with $\sum m_\nu = 0.1$ and 0.3~eV, whose summed difference is 0.02~eV larger than in our analysis (we compare $\sum m_\nu = 0.06$ to $0.24$). They too find a suppression of $20-30\%$ in the $\kappa = 0.1-0.2$ domain. Finally, \citet{fong2019impact} studied the impact of massive neutrinos on WL peak counts in the BAHAMAS simulation. Amongst others, they consider models with $\sum m_\nu = 0.06$ and 0.24~eV. They see a difference of $10-20\%$ up to SNR\,=\,9, slightly smaller than the difference we find.

The comparisons in this section show that the number densities of peaks are highly sensitive to cosmology. Variations of a few per cent in $\Omega_\mathrm{m}$ or $\sigma_\mathrm{8}$ cause changes in the WL peak abundances of $\approx 20-60\%$. The differences can be understood qualitatively by considering the impact of cosmological parameters on the HMF.

\subsection{Separability of cosmological and hydrodynamical effects}\label{sec:hydrosep}
\begin{figure}
	\includegraphics[width=\columnwidth]{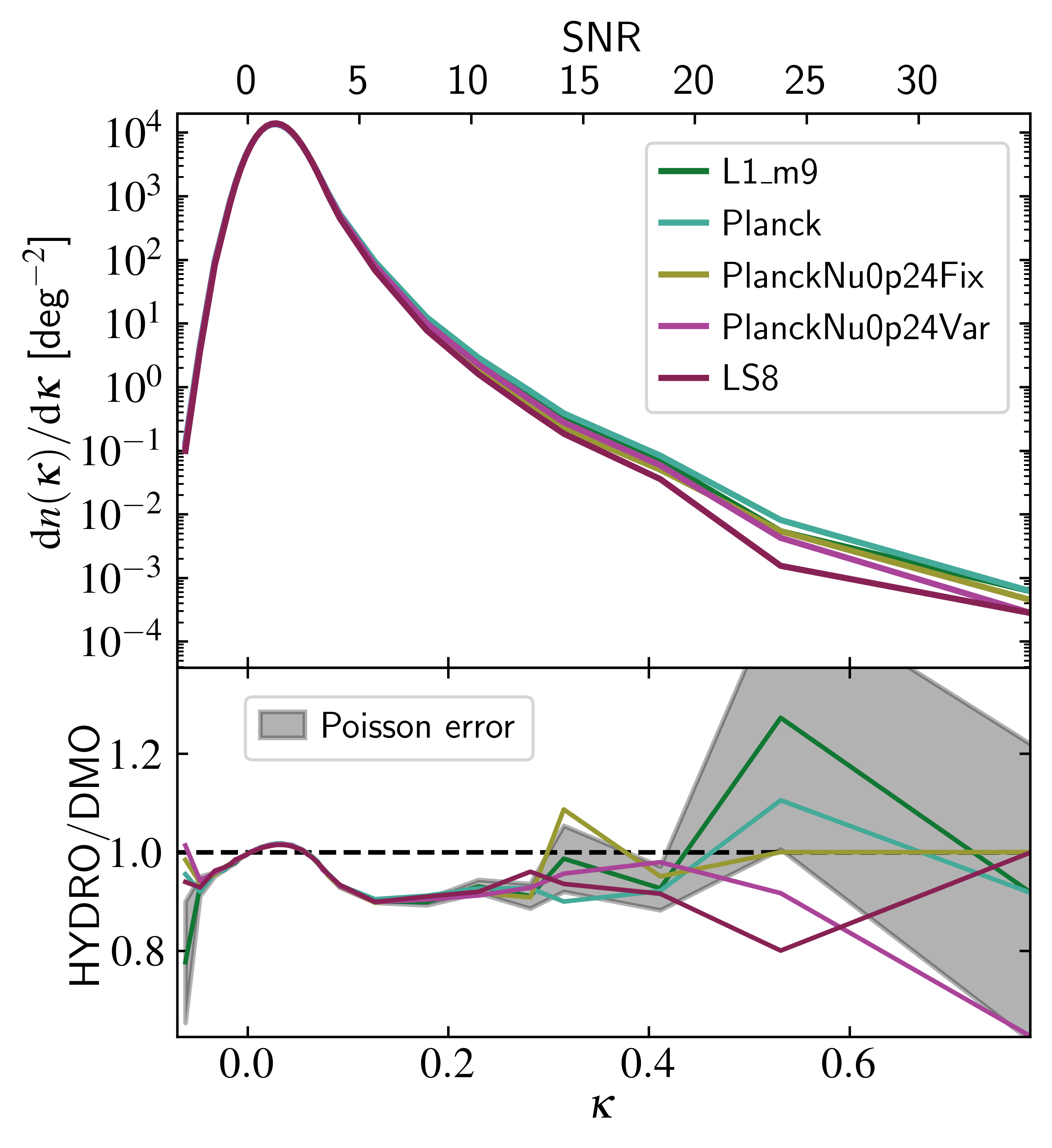}
    \caption{As Fig.~\ref{fig:dmo_vs_hydro} but comparing the hydrodynamical runs with different cosmologies but the same baryonic calibration to their corresponding DMO runs. Whereas the variation between the DMO signals in Fig.~\ref{fig:cosmo_vars} is up to 300\%, the baryonic suppression (as seen in the bottom panel) is similar for all cosmology variations, illustrating the separability of the cosmology and baryonic impact.}
    \label{fig:cosmo_hydro_sep}
\end{figure}

Finally, we explore the separability of the cosmological and astrophysical impact on the number density of WL peaks. In Fig.~\ref{fig:cosmo_hydro_sep}, the top panel shows the number density of WL peaks for the cosmology variations, which were done at fixed baryonic calibration, in the full-hydrodynamical runs. The bottom panel shows, for each cosmology variation separately, the ratio to its corresponding DMO run. Despite the big differences seen between the cosmology variations in Fig.~\ref{fig:cosmo_vars}, when comparing the variations to their own DMO run, the baryonic suppression is very similar for all cosmology variations, as shown in the bottom panel of Fig.~\ref{fig:cosmo_hydro_sep}, illustrating that the cosmological and hydrodynamical effect are largely separable. The deviation between the number densities of WL peaks up to $\kappa = 0.2$ (0.4) is not more than 1 (10)~\%, and generally falls within the Poisson noise for L1$\_$m9 as indicated by the gray-shaded area, whereas the cosmology variations in Fig.~\ref{fig:cosmo_vars} show up to an order of magnitude larger differences in the same $\kappa$ regime.

We can understand the separability from the different impacts of the two processes on halos. We have seen that the cosmology dependence translates itself into changes in the total number of halos at fixed mass (through $\Omega_\mathrm{m}$ and $\sigma_8$) and their halo mass (through mass being carried away by neutrinos), whereas the baryonic physics displaces gas and removes mass from the centers of halos. As these processes are physically independent we expect to see the cosmology and baryonic separability for the number density of WL peaks. The comparison shows that the two processes are indeed separable, which demonstrates the potential to model the impact of cosmology on WL peaks based on DMO simulations only.

\section{Discussion}\label{ch:discussion}
We start the discussion by comparing the different sets of baryonic and cosmological variations of the previous sections. First, we compare the impact of all baryonic variations to the difference in suppression between the fiducial DMO and hydrodynamical simulation. Fig.~\ref{fig:dmo_vs_hydro} shows that the fiducial baryonic suppression is $\approx 10\%$ for $\kappa = 0.1$ and that the suppression decreases for larger WL convergence values. Comparing this to the suppression and enhancement of the different feedback variations in Fig.~\ref{fig:gas_fracs} and~\ref{fig:feedback_vars}, we see that the other baryonic feedback variations show a suppression of 5-30\% compared to the DMO signal, for $\kappa = [0.1,0.4]$. 

Comparing the impact of cosmology and baryonic physics, we can study the degeneracy of the two effects. With the limited number of variations that we consider in this research, we do not find clear evidence that the two effects are non-degenerate, meaning that the two effects impact the same $\kappa$ ranges similarly, and the two processes each have to be modeled carefully. However, we stress that we only have a limited number of variations and that to properly address this question one should compare variations that jointly vary in cosmology and baryonic physics, which we aim to do in the future. Additionally, including other observables may help break the degeneracy.

The variations between the fiducial and the Planck or LS8 models, as shown in Fig.~\ref{fig:cosmo_vars} are larger than the differences between the runs that vary the gas fractions, as seen in Fig.~\ref{fig:gas_fracs}, which span 10$\sigma$ in gas fractions. The enhancement in the number density of the Planck cosmology is about twice as large as for the fgas$+2\sigma$ model and LS8 suppresses the number density of peaks $2-3$ times as much as the fgas$-8\sigma$ model. These differences illustrate that within the FLAMINGO simulation suite, the number density of peaks is more sensitive to reasonable differences in cosmological parameters than to the broad range of explored gas fractions, some of which are already ruled out by present observations of galaxy clusters (assuming there are no unrecognized systematics). It is important to keep in mind that even among models that reproduce the observed galaxy stellar mass function and predict the same cluster gas fractions, the exact effect of the baryonic feedback can depend on the adopted subgrid model, as we saw in the comparison of the jet to thermal AGN feedback models. Other parts of the astrophysics model could also play an important role. However, since the covered range in gas fractions is large, the analysis still highlights that the number density of peaks is more sensitive to the total amount of matter, set by $\Omega_\mathrm{m}$, and its clumpiness, as set by $S_8$, than to the uncertainty in the distribution of gas in the Universe, resulting from baryonic feedback. Nevertheless, the astrophysical uncertainties are not negligible compared to the effect of varying the cosmology.

Next, we focus on the different impacts of neutrinos and baryons. To this end, we compare the relative difference of the gas fraction models to the fiducial model (Fig.~\ref{fig:gas_fracs}) and the heavier neutrino models to the Planck model (Fig.~\ref{fig:cosmo_vars}), as the heavier neutrino models are variations on the latter model. The suppression by the neutrino models with $\sum m_\nu = 0.24$~eV in the intermediate convergence regime ($\kappa \approx [0.1,0.4]$) is slightly stronger than the difference between fgas$+2\sigma$ and fgas$-8\sigma$. Our analysis thus shows that WL peak counts are more sensitive to the difference in cosmological parameters between the two sets of cosmologies currently favored by either CMB or WL measurements than the uncertainties in the impact of SN and AGN feedback leading to different cluster gas fractions or galaxy stellar mass functions.

While the cosmology variations cause larger differences in the number density of peaks, it is clear that the uncertainty on WL peaks arising from baryonic effects is not negligible for a Stage IV inference. Both the cosmology and baryonic variations considered in this paper are on the extreme side in the sense that we expect Stage IV WL surveys to be able to discriminate between them, which means that the baryons have to be properly modeled. In particular, the impact of baryons and cosmology shows up in the same convergence regime with similar behavior, but Section~\ref{sec:hydrosep} has illustrated that the cosmology and baryonic impact are mostly separable in the sense that the baryonic suppression is insensitive to the cosmology. The uncertainties due to baryons may be overcome by considering a joint inference of 2-point and non-Gaussian statistics. For example, \citet{Semboloni2013} have shown that baryonic feedback impacts two- and three-point shear statistics differently. Using a combination of different WL statistics, the uncertainties due to baryons could be calibrated simultaneously whilst constraining the cosmology. 

Finally, although the low-$\kappa$ regime ($\kappa < 0.1$) is strongly affected by smoothing and noise, there are still clear differences between the simulations. For instance, whereas LS8 has a suppressed number density of WL peaks at the intermediate- and high-$\kappa$ end, it has an increased number of peaks around $\kappa = 0.05$ (Fig.~\ref{fig:cosmo_vars}). Although the difference is only a few per cent, this regime contains several orders of magnitude more peaks, and therefore the impact of Poisson noise and cosmic variance (Fig.~\ref{fig:dmo_vs_hydro}) vanishes, and it may potentially help in discriminating between the models. However, the interpretation of these peaks is more difficult because the systematics affecting this regime need to be very well understood. Also, the physical interpretation of these peaks is less straightforward as they are caused by multiple objects along the line-of-sight \citep[][]{yang2011cosmological}{}{}, and we note that the baryonic impact and cosmology variations are of similar magnitude around $\kappa = 0.05$, whereas the variations are relatively greater for the cosmology variations in the larger kappa regime. We aim to explore both the constraining power and the physical origin of these peaks in future research.

\section{Summary and Conclusions}\label{ch:conclusion}
We have studied the impact of baryonic feedback on the distribution of peaks in weak lensing convergence maps. We used the state-of-the-art cosmological hydrodynamical FLAMINGO simulation suite, in which the parameters of the subgrid prescription for feedback are calibrated to match the observed $z=0$ galaxy stellar mass function and the gas fraction in low$-z$ clusters. To mimic the signal of a Stage IV weak lensing survey, we used a full-sky ray-tracing method with a \textit{Euclid}-like source redshift distribution and shape measurement noise. We studied numerical convergence (Fig.~\ref{fig:mass_res}) and cosmic variance (Fig.~\ref{fig:cosmic_var}), showing that both are under control in the intermediate weak lensing convergence regime ($\kappa \approx [0.1,0.4]$), where most cosmological information is contained. Our analysis shows that for $\kappa \gtrsim 0.4$, the signal is dominated by Poisson noise and cosmic variance, which indicates a limit to the usefulness of high-$\kappa$ WL peaks as all WL surveys will suffer from these effects. 

Our results agree with previous studies \citep[][]{coulton2020weak,Osato2021baryons,FerlitoMTNGconvergence}{}{} as we find that baryonic feedback processes, which expel gas from the centers of haloes and thereby decrease their mass, lead to a suppression of the number density of WL peaks of $\sim$10\% in the cosmologically informative WL convergence regime (Fig.~\ref{fig:dmo_vs_hydro}).

At fixed cosmology, initial conditions, and observer position, we compared 9 models that differ in their galaxy stellar mass function, cluster gas fraction, and/or AGN feedback implementation; we find baryonic suppressions in the range 5-30\% (Fig.~\ref{fig:gas_fracs}~\&~\ref{fig:feedback_vars}). Our results show a clear trend with cluster gas fraction; in simulations with a lower gas fraction, the number density of peaks is more suppressed. The differences may be understood from an individual halo perspective: cluster haloes with lower gas fractions have lost more gas from their center, resulting in a lower mass and hence a suppression of the WL signal and number density of peaks. Similarly, calibrating the feedback prescription so as to reduce the amplitude of the galaxy stellar mass function also lowers the number density of WL peaks in the well-resolved convergence regime ($\kappa \approx [0.1,0.4]$). The differences between the thermal and kinetic jet AGN feedback models, calibrated to obtain the same galaxy mass function and cluster gas fractions, illustrate that the baryonic suppression is not fully specified by these two observables but also depends on the details of the feedback implementation. 

We compared simulations with different values of the cosmological parameters, favored either by the CMB or by WL measurements, and with different neutrino masses. The impact of these variations can be understood by reference to the effect that the cosmological parameters have on the HMF. The cosmology variations show greater differences than the baryonic variations at fixed cosmology, suppressing (enhancing) the number density of WL peaks up to 60 (20)\% (Fig.~\ref{fig:cosmo_vars}).

Our results illustrate that WL peak counts are a useful statistic to constrain cosmological parameters with upcoming WL surveys. Within the FLAMINGO simulation suite, we have shown that variations in cosmology have a larger impact than variations in the treatment of baryonic physics. However, the differences induced by baryons are larger than the expected accuracy of upcoming surveys and must therefore be modeled carefully. Importantly, our analysis shows that while the effects of baryonic physics are well behaved and can be understood, their effects are qualitatively similar to those due to variations in cosmology. However, the impact of baryons is insensitive to the changes in cosmology that we explored, suggesting that the effect of cosmology can be investigated using DMO simulations.\footnote{Note in particular that large variations in $\Omega_\mathrm{b}/\Omega_\mathrm{m}$ have not been explored in this study.} The analysis shows that the deviation of baryonic suppression at different cosmologies up to $\kappa = 0.2$ (0.4) does not exceed 1 (10)\%  (Fig.~\ref{fig:cosmo_hydro_sep}) and that the $\kappa > 0.4$ regime is dominated by Poisson noise and cosmic variance.

Non-Gaussian statistics could be used to disentangle baryonic from cosmological effects. In order to exploit WL peaks or other beyond-Gaussian statistics, both neutrinos and baryons need to be carefully modeled. Multiple WL statistics could be used to simultaneously calibrate the baryonic physics and to constrain the cosmology. Additionally, it will be necessary to forward model photometric redshifts, intrinsic alignments, masks, multiplicative shear bias as well as tomographical analyses \citep[e.g.][]{Fluri2018,Zurcher2021,Zhang2022}{}{}. Ideally, the simulations will cover a broad enough range of cosmological and astrophysical feedback parameters. In the future, such approaches will become possible by using emulators built on top of simulation suites in which the feedback implementation and the cosmological parameters are varied simultaneously.

\section*{Acknowledgements}
JCB thanks Christopher Davies for the helpful discussions on peaks in HEALPix maps. CSF acknowledges support by the European Research
Council (ERC) through Advanced Investigator grant, DMIDAS (GA
786910). This work is partly funded by research programme Athena 184.034.002 from the Dutch Research Council (NWO). This work used the DiRAC@Durham facility managed by the Institute for Computational Cosmology on behalf of the STFC DiRAC HPC Facility (\url{www.dirac.ac.uk}). The equipment was funded by BEIS capital funding via STFC capital grants ST/K00042X/1, ST/P002293/1, ST/R002371/1 and ST/S002502/1, Durham University and STFC operations grant ST/R000832/1. DiRAC is part of the National e-Infrastructure.

\section*{Data Availability}
The data supporting the plots within this article are available on
reasonable request to the corresponding author. The
FLAMINGO simulation data will eventually be made publicly available. 



\bibliographystyle{mnras}
\bibliography{example} 




\appendix

\section{Box rotation}\label{app:box_rotation}
In this Appendix, we explore the impact of different box rotation strategies. As all the baryonic feedback variations were carried out in 1\,cGpc boxes, which need to be replicated 6 times to cover the distance to $z=3$, the approximate redshift range necessary for a Stage IV WL survey (see Fig.~\ref{fig:source_dist} and the green arrow indicating the L1 box length). As the lightcone extends to $z=3$ in all directions, in total, 13 L1 box replications are necessary for the L1 runs.

To test the impact of the replication along the line of sight and the strategy to deal with the shell rotations, we compare the signal of the L1$\_$m10$\_$DMO run to that of the L5p6$\_$m10$\_$DMO run. This simulation, which contains 5040$^3$ DM and $2800^3$ neutrino particles, was run in a 5.6\,Gpc box at the fiducial cosmology and has the same resolution as the low-resolution (m10) L1 run. For this cosmology, 5.6\,cGpc corresponds to a redshift of $z=2.2$, which covers the largest part of the redshift range as indicated in Fig.~\ref{fig:source_dist}. However, as the lightcone extends in all directions, the independent volume extends only to $L=2.8$~cGpc ($z=0.8$), which corresponds approximately to the mean of the redshift distribution we use (see Fig.~\ref{fig:source_dist}). Here, to make sure we do not repeat over any structures in the 5.6\,cGpc box, we adopt a delta-function source redshift distribution, $n(z) = \delta(z=0.8)$. 

We then compare L5p6$\_$m10$\_$DMO with L1$\_$m10$\_$DMO for different treatments of the redshift shells, where we consider the signal from the 5.6\,cGpc run to be the ground truth. For L1$\_$m10$\_$DMO we test three different rotation prescriptions. First, we apply no rotation to any of the mass shells. In this case, depending on the angular position on the sky, we sum up to 6 times over the same structure (though at different stages of evolution) along the line of sight. Second, we rotate every shell randomly to prevent running into the same structure. Third, we randomly rotate the shells whenever the diameter of the lightcone is larger than the box length. In this way, we still minimize the number of repetitions but also do not superfluously erase any possible correlations along the line of sight.

\begin{figure}
	\includegraphics[width=\columnwidth]{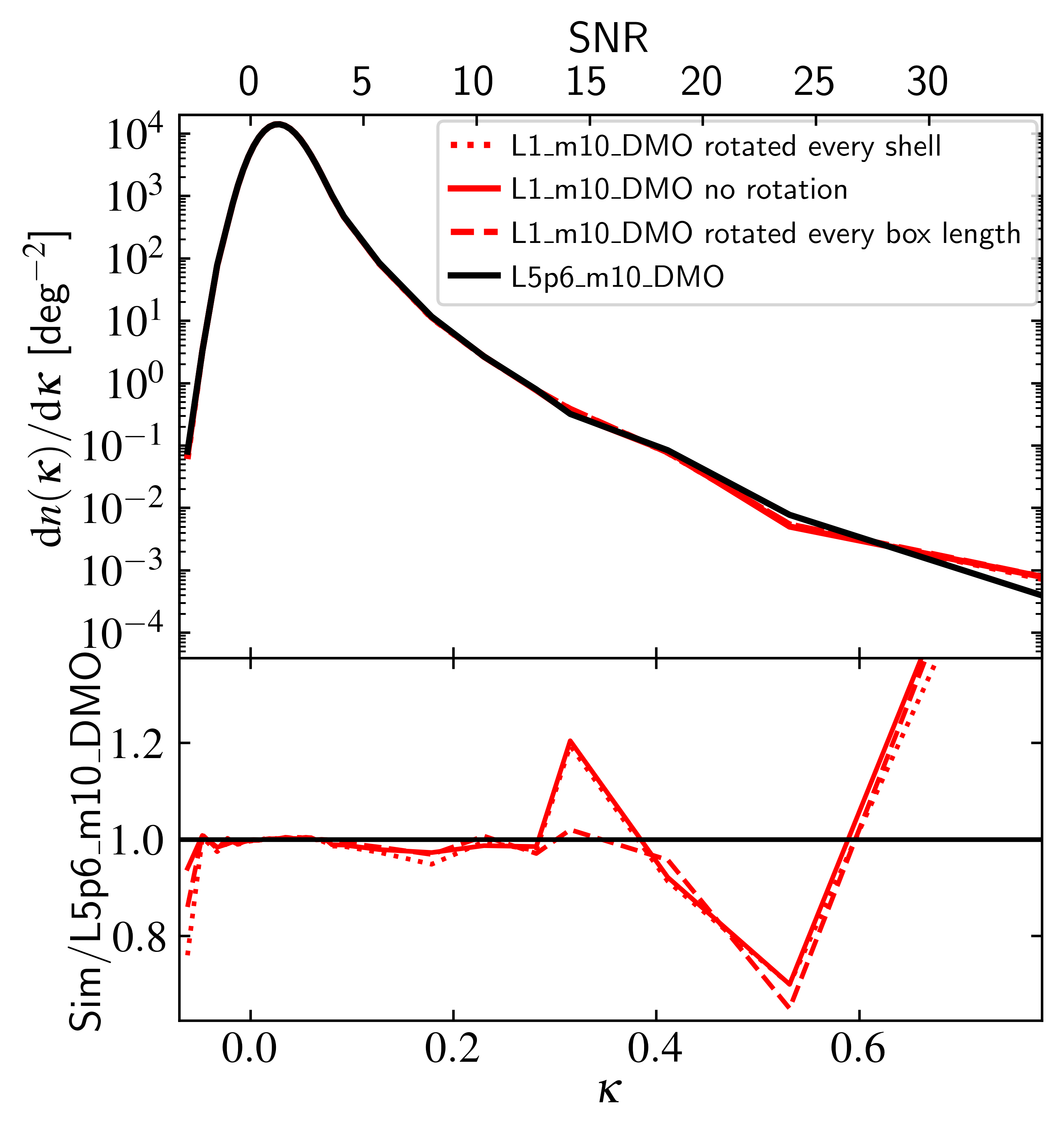}
    \caption{Top: number densities of WL peaks for the L5p6$\_$m10$\_$DMO model (black) and L1$\_$m10$\_$DMO run (red) with different mass shell rotation strategies where the shells in the latter simulation are not rotated (dashed), every shell is randomly rotated (dotted) or the shells are randomly rotated every half box length (solid). Bottom: ratio to the L5p6$\_$m10$\_$DMO signal. The different rotation methods only show minor differences.}
    \label{fig:box_rotation}
\end{figure}

Fig.~\ref{fig:box_rotation} shows the number densities of peaks for the different methodologies. The L1$\_$m10$\_$DMO variants are indicated in red where the no rotation, rotation of every shell, and rotation of every half box length are indicated by the solid, dotted, and dashed lines, respectively. The bottom panel compares the three rotation strategies to the signal measured in the L5p6$\_$m10$\_$DMO run, which corresponds to the black curve in the top panel. As the different box sizes are different realizations of the same Universe, we do not expect them to show perfect agreement but be impacted by cosmic variance. As is clear from the bottom panel, the variations in replication strategy have only a minor impact on the measured signal. This agrees with \citet{Upadhye2023}, who studied similar rotation strategies for the CMB lensing angular power spectrum. All three rotations show some minor suppression compared to the L5p6 box, independent of the rotation strategy. Only at $\kappa \approx 0.3$, there is a difference between the rotation strategies. Here the not rotated and rotated every shell procedures give a 20\% difference while the rotated every half box length method yields almost perfect agreement with L5p6$\_$m10$\_$DMO. In our analysis, we choose to randomly rotate every half box length and we apply the same rotation to simulations of the same box size.

\section{Smoothing and Noise}\label{app:smooth_noise}
\begin{figure}
	\includegraphics[width=\columnwidth]{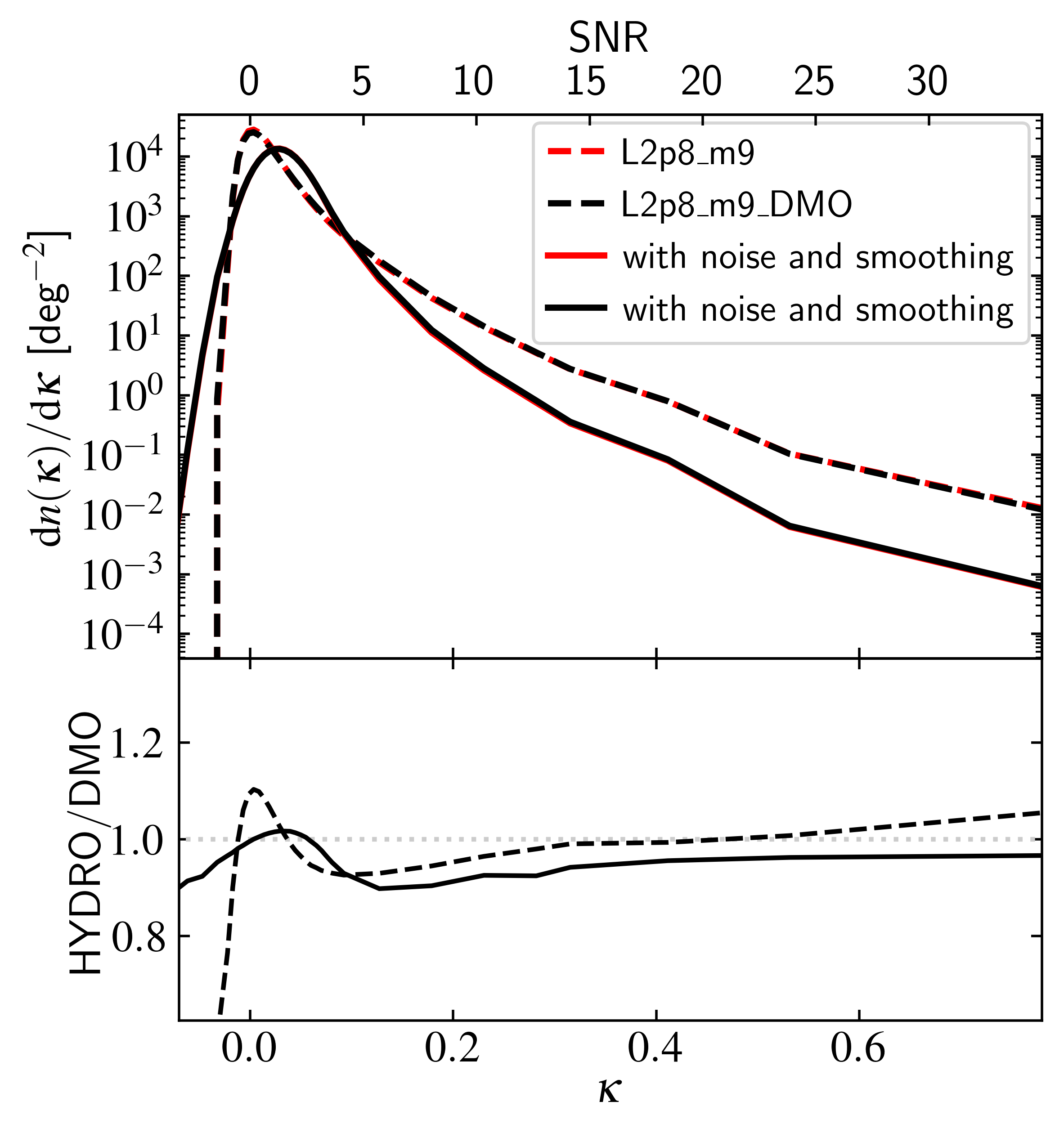}
    \caption{Top: number densities of WL peaks for the mean of the 8 observers in the L2p8$\_$m9 (red) and  L2p8$\_$m9$\_$DMO (black) runs with different treatment for the applied smoothing and noise. The solid and dashed lines correspond to number densities measured from maps that have no smoothing or noise, or both noise and smoothing, respectively. The lines with the same linestyle are directly behind each other.  The bottom panel shows the ratio of the hydrodynamical to DMO number densities for maps that have had the same noise and smoothing treatment. The entire WL convergence regime is impacted by the smoothing and noise, indicating a careful analysis has to be done when carrying out an inference with real observations.}
    \label{fig:smoothing_noise}
\end{figure}

In this Appendix, we illustrate the effect of the applied galaxy shape noise and smoothing on the number densities of WL convergence peaks. The noise and smoothing are applied to mimic the expected observed galaxy shape noise (Equation~\ref{eqn:noise}) and the effect of any instrument carrying out a WL survey. We compare distributions without noise and smoothing to those with noise and smoothing in Fig.~\ref{fig:smoothing_noise}. If smoothing is applied, the map is smoothed with a Gaussian kernel with FWHM = 1 arcmin. The figure includes the mean of the 8 virtual observers in the hydrodynamical L2p8 (red) and L2p8$\_$DMO (black) runs. The solid and dotted lines correspond to instances in which the final convergence map contains no smoothing and noise, or both noise and smoothing, respectively. The red curves are directly behind their corresponding black curves.

The bottom panel in Fig.~\ref{fig:smoothing_noise} gives the ratio of the hydrodynamical to DMO number densities as measured for the maps that used the same noise and smoothing treatment. Depending on the convergence regime, the noise and smoothing have a different impact. For $\kappa \approx 0$, the smoothing and noise wash out the largest baryonic effects, as can be seen in the bottom panel where the difference at $\kappa = 0$ is suppressed from 10\% to 2\%. The ratio of the maps without any treatment shows an enhancement of peaks with $\kappa \approx 0$ in the hydrodynamical run, which has almost completely vanished after the smoothing and noise are applied. For $\kappa > 0.1$, the number of peaks is suppressed significantly by the map treatment, as can be seen by comparing the dashed to solid lines in the top panel. The smoothing primarily impacts the total number of peaks and the high-$\kappa$ end as it levels out fluctuations on the smallest scales and lowers the values of the highest peaks, whereas the noise is primarily responsible for erasing the difference between the runs around $\kappa = 0$. The comparison illustrates that the entire $\kappa$ regime is sensitive to the treatment of the WL convergence maps, which makes sense as smoothing the maps effectively smooths out overdensities, but it does illustrate that when comparing to observations a more dedicated study into these systematics should be carried out.

\bsp	
\label{lastpage}
\end{document}